\documentclass[%
 aip,
 amsmath,amssymb,
  reprint,%
jcp
]{revtex4-1}

\usepackage{graphicx}
\usepackage{dcolumn}
\usepackage{bm}

\usepackage[utf8]{inputenc}
\usepackage[T1]{fontenc}
\usepackage{mathptmx}

\usepackage{xcolor}
\usepackage{cancel}
\usepackage{soul}

\begin{document}


\title[]{Coarse-graining strategy for modeling effective, highly diffusive fluids with reduced polydispersity: A dynamical study}

 \author{Thomas Heinemann}
 \email{thomas.heinemann@alumni.tu-berlin.de}
\affiliation{Department of Chemistry, Seoul National University, Seoul 08826, Korea  }
\author{YounJoon Jung}
\email{yjjung@snu.ac.kr}
\affiliation{Department of Chemistry, Seoul National University, Seoul 08826, Korea  }


\date{\today}

\begin{abstract}
We present a coarse-graining strategy for reducing the number of particle species in mixtures to achieve a simpler system with higher diffusion
while preserving the total particle number and characteristic dynamic features.
As a system of application, we chose the bidisperse Lennard-Jones-like mixture discovered by Kob and
Andersen [Phys. Rev. Lett. \textbf{73}, 1376 (1994)] possessing a slow dynamics due to the fluid's multi-component character with its apparently unconventional choice for the pair potential of the type-A--type-B arrangement. 
We further established in 
a so-formed coarse-grained and temperature-independent
monodisperse system 
an equilibrium structure with a radial distribution function resembling its mixture counterpart.
This one-component system further possesses similar dynamic features like glass transition temperature and critical exponents
while subjected to Newtonian mechanics.
This strategy may finally lead to the manufacturing of new nanoparticle/colloidal fluids by experimentally modeling only the outcoming effective pair potential(s) and no other macroscopic quantity.
\end{abstract}

\maketitle

\section{\label{sec:Introduction}Introduction}

In this work we introduce a novel type of coarse-graining procedure for isotropic, polydisperse, Newtonian systems for 
achieving an effective system with fewer components possessing higher diffusion while the total particle number, the Newtonian mechanics, as well as relevant characteristic dynamic features are preserved.
By coarse-graining (CG), we understand the systematic treatment of microscopic details and dynamics from a coarser perspective.
Its purpose is to achieve a simpler system as well as having enlarged length or time scales in computer simulations by effectively considering microscopic details.
Common CG methods in the literature include force matching schemes,\cite{Loewen1993,Ercolessi1994,Izvekov2005,Izvekov2005jcp}  the relative entropy method,\cite{Shell2008} the conditional reversible work method,\cite{Brini2011}  the inverse Monte Carlo method,\cite{Lyubartsev1995,Lyubartsev1997} the iterative Boltzmann inversion method\cite{Soper1996,MullerPlathe2002} or hybrid schemes.\cite{Ruhle2011}
These methods are primarily used to develop effective force fields among a small set of coordinates, called reaction coordinates\cite{Heinemann2014,Heinemann2017} or collective variables, that form projections of microscopic degrees of freedom.
A frequent choice of such coordinates is given by the molecules' center-of-mass positions.\cite{Likos2001}

Here, our presented coarse-graining strategy differs from the previously described standard strategy since we aim at effectively reducing the number of species while not reducing the particles, i.e. the number of collective variables equals the 
number of microscopic degrees of freedom.
We recognized that top-down coarse-graining involving species reduction has been performed already.\cite{ebrahimi2016effect,Peters2018}
We, on the contrary, perform coarse-graining in a bottom-up philosophy and thereby achieve a systematic reduction or even a neglect of multi-component character, which is regarded as microscopic detail. 
By eliminating such detail, one can achieve a higher diffusion and accordingly larger length and time scales in \textit{in-silico} simulations as a consequence of a smoother energy landscape  (see Ref.~\citenum{Zwanzig1988}).
A higher diffusion might lead to an improved wetting behavior as required in real-world applications like motor oil, hot glue, or inks.
Besides better wetting, also the friction can be reduced with higher diffusion as required in machinery, e.g., via a single-component nanoparticle oil adhesion aiming to imitate a multi-component oil adhesion which, from experience, tends to be less tribological.\cite{Shahnazar2016}
However, increasing diffusion might also lead to a change of relevant characteristic dynamic features like vitrification and glass transition.
Preserving vitrification properties might be even considered as a necessity, e.g., in hot glue and ink-like applications.
A common solution for controlling the dynamics, and thus preserving dynamic features, in an effective particle system is to modify the equations of motion with stochastic terms (e.g., Langevin dynamics, dissipative particle dynamics,\cite{Hoogerbrugge1992, Koelman1993, Espanol1995} \dots ),
i.e., density fluctuations stemming from a multi-component character could be emulated by density fluctuations from the reduced species set. 
We, on the contrary, stay with the Newtonian mechanics also in the CG system and do not intend to make the dynamics more complex, i.e. in our microscopic as well as our CG system we consider all particles to obey the Newtonian equations of motion (besides a weak temperature coupling).
In that way, we aim to get all targeted dynamic aspects by having a proper choice for the effective pair potential and take possible structural deviations into account.
From the experimental point of view, one can, thus, focus on how to imitate that pair potential.

With respect to the application point of view, there is also a direct demand for reducing
polydispersity in melts towards one single component as long as the effective pair potential is not too complex for manufacturing.
Potential examples are given by single-component polymer gels,\cite{Daniel2016} which do not require a solvent and thereby avoid unwanted species separations, 
or sealants,\cite{Chronister2000} which do not have to be in a de-mixed state before being used at high temperatures. 

\begin{figure}
\begin{center}
 \includegraphics[width=0.85\columnwidth]{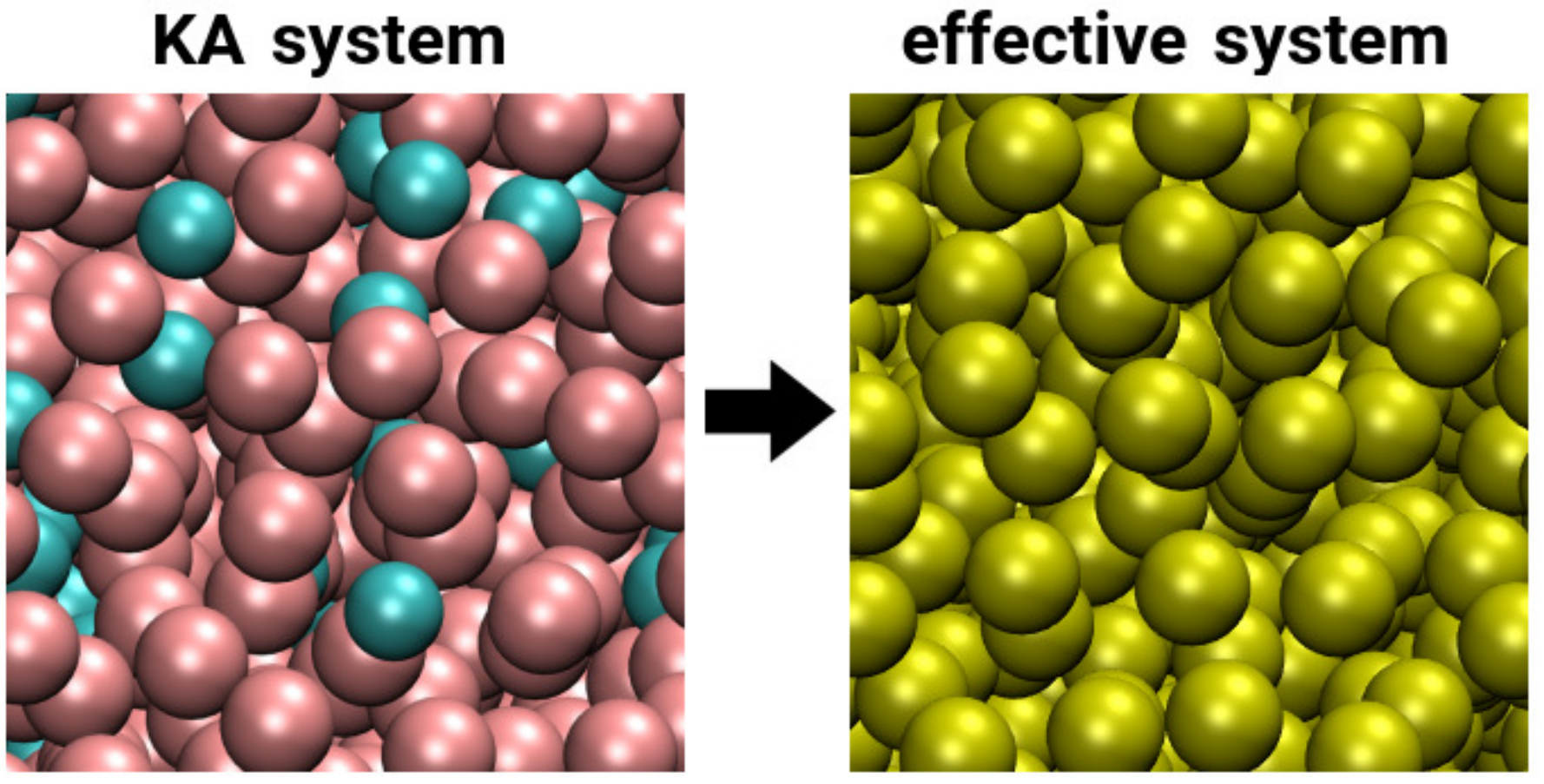}  
\end{center}
 \caption{Transition from the bidisperse Kob-Andersen fluid (particle species A=red, B=turquoise) towards an effective monodisperse fluid (yellow particles).}
 \label{fig:ibischemepics}
\end{figure}

To test our method, we chose the Kob-Andersen (KA) mixture,\cite{Kob1994,Kob1995} which is a bidisperse Lennard-Jones mixture with glass-forming ability whose Lennard-Jones potential range and strength parameters do not fulfill the common Lorentz-Berthelot mixing rule.\cite{Boda2008}
The KA mixture, when supercooled but still being warmer than its glass transition temperature, has a very long-lasting isotropic regime before it crystallizes.
Mixtures with similar dynamic features also exist for other models,\cite{Pedersen2010,Kim2013} even in different dimensions.\cite{Bruning2008,Flenner2015}
By creating a coarse-grained or effective KA system consisting of a one-particle species, as schematically shown in Fig.~\ref{fig:ibischemepics}, 
a higher diffusion and lower relaxation time may follow.
These properties, however, act negatively on the glass-forming ability or influence its critical temperature due to less/no entropy of mixing.\cite{pinal2008entropy}
Therefore, the question to be investigated is: Can we yield a monodisperse glass-former resembling in structure but possessing a higher diffusion than the KA mixture?
Prior model approaches for monodisperse glass-formers include particle shape anisotropy,\cite{Pfleiderer2008} pinned particles,\cite{Bhowmik2006} ultrasoft interactions,\cite{Ikeda2011}
an attractive pair potential with a shoulder\cite{Dzugutov1992}
or the use of double well\cite{VanHoang2008} as well as undulating pair potentials.\cite{Elenius2010} 
As we see later, our CG approach also leads to undulating pair potentials.

In order to further specify our polydispersity reduction, we require that the all-particle radial distribution function (RDF), which does not distinguish between the particle type, is set to remain unchanged.
During the course of this study, we will slightly mitigate this requirement.
As the principal RDF-matching method, we chose the iterative Boltzmann inversion (IBI) method\cite{Soper1996} which, when converged, leads to an effective model entirely build on an effective pair potential.
In contrast to other CG approaches aiming at conserving dynamic features,\cite{Izvekov2006,Davtyan2015} the IBI method conserves with the RDF an equilibrium quantity.
Thus, it is interesting to scrutinize if, or under which modifications, the here presented CG method is still able to maintain the ability of the so-formed monodisperse system to appear in a glass phase.
Within the dynamic context, the stability of the isotropic phase is another aspect we set out to scrutinize since monodisperse systems tend to have shorter relaxation times and crystallize at a fast pace\cite{narumi2007different,Tokuyama2011} when supercooled.
Even though the particles' kinetic energies seem quite low for establishing a fast crystallization, a large diffusion, however, could also mean that crystallization barriers could be overcome easier.
As we later point out in Sec.~\ref{sec:Assessing glassy aspects of the systems' dynamics}, we could also imagine ways of moderately reducing the high gain in diffusion in our effective model at the cost of a larger structural mismatch.

The remainder of this article is organized as follows. In Sec. \ref{sec:RDF-matching coarse-graining approach}, we present the theory of the RDF-matching CG approach by describing the transformation from the bidisperse KA mixture towards an effective monodisperse fluid.
Then in Sec.~\ref{sec:Temperature-independent RDF-matching approach}, we  propose temperature-independent versions of this approach by introducing the models M0.9 and M0.45.
Later on, we present in Sec.~\ref{sec:Assessing glassy aspects of the systems' dynamics}
an investigation of the dynamics covering the mixture and systems created through both temperature-independent models.
Finally, we conclude our findings in Sec. \ref{sec:Conclusions}.

\section{\label{sec:RDF-matching coarse-graining approach}RDF-matching coarse-graining approach}
We motivate in the following the RDF-matching coarse-graining (CG) approach using the Kob-Andersen (KA) mixture\cite{Kob1994, Kob1995} in Subsec.~\ref{sec:Tranforming the KA-mixture into a monodisperse fluid} and perform initial assessments on the outcoming effective monodisperse
fluid in Subsec.~\ref{sec:Initial assessments of the RDF-matching approach}. In these assessments we point out the method's limitations, which we set to overcome later in Sec.~\ref{sec:Temperature-independent RDF-matching approach}.

\subsection{\label{sec:Tranforming the KA-mixture into a monodisperse fluid}Tranforming the KA-mixture into a monodisperse fluid}

\begin{figure}
\begin{center}
 \includegraphics[width=0.95\columnwidth]{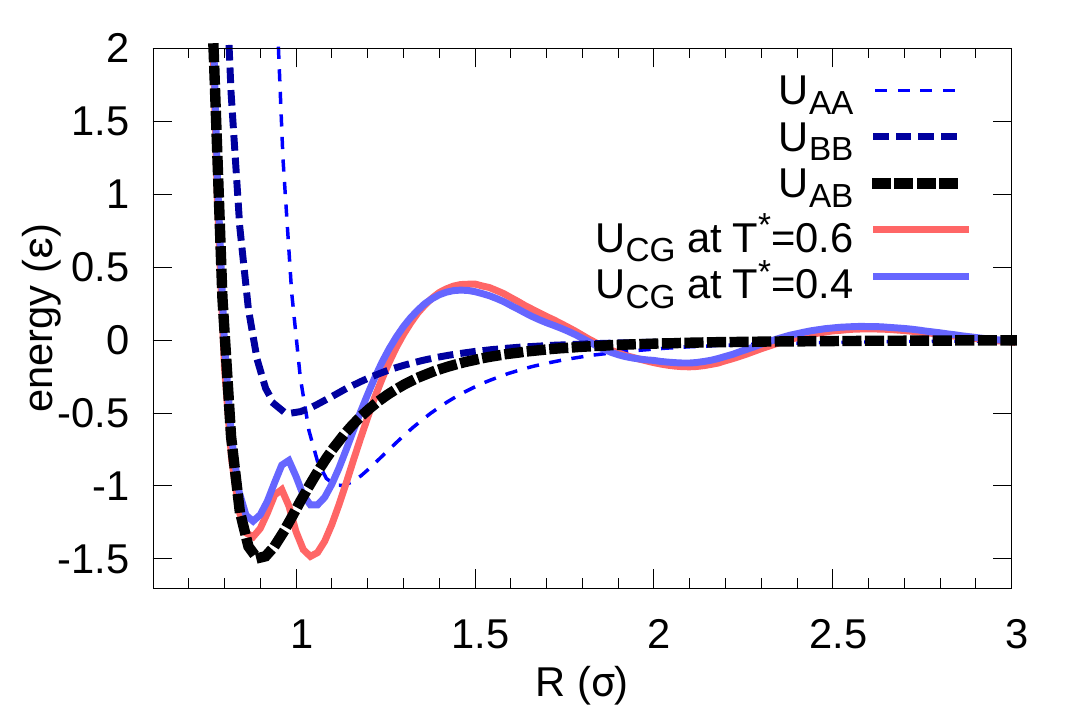}  
\end{center}
 \caption{Kob-Andersen pair potentials for AA, BB, and AB particle pairs, and our effective potential at $T^*=k_{\rm B} T/\epsilon=0.4, 0.6$.}
 \label{fig:uofrkobandersenwithueff}
\end{figure}

The KA mixture consists of Lennard-Jones particles of mass $m_p$ comprising two different species, labeled with A and B, whereas there are four times more A than B particles.
It is these few B particles which prevent the other particles, that form the $80\%$ majority, from quickly crystallizing.
The set of pair potentials as a function of the inter-particle distance $R$ in that mixture is defined through
\begin{multline}
 U_{\alpha \beta}(R)=\\
 =\begin{cases}
                      U_{\rm LJ}(R,\epsilon_{\alpha \beta},\sigma_{\alpha \beta})\! -\! U_{\rm LJ}(2.5 \sigma_{\alpha \beta},\epsilon_{\alpha \beta},\sigma_{\alpha \beta}) &\!\!\!\!, R\le 2.5\! \sigma_{\alpha \beta}\\
                      0 &\!\!\!\!, {\rm otherwise}
                     \end{cases}
\end{multline}
with $\alpha,\beta \in \{A,B\}$, and the Lennard-Jones parameters for the well depth
\[
 \epsilon_{\rm AA}= \epsilon \qquad \epsilon_{\rm AB}=1.5 \epsilon \qquad \epsilon_{\rm BB}=0.5 \epsilon  \]
 and contact distance
 \[
 \sigma_{\rm AA}= \sigma \qquad \sigma_{\rm AB}=0.8 \sigma \qquad \sigma_{\rm BB}=0.88 \sigma\text{.}
\]
Corresponding potential curves are depicted in Fig.~\ref{fig:uofrkobandersenwithueff} as dashed lines.
In accordance with the original KA system, we consider in the effective monodisperse system the same  particle density $\rho$ being fixed at $\rho\sigma^3=1.2$. 
The first trial within our coarse-graining approach consists in identifying at each considered temperature the all-particle RDF (denoted with $g$) of the bidisperse KA system with the RDF of the coarse-grained (CG) or effective monodisperse
system, i.e.
\begin{align}
g_{\rm KA}\stackrel{!}{=} g_{\rm CG}.
\end{align}
This is equivalent to identifying the mean force profiles.\cite{Kirkwood1935}
The definition of the all-particle RDF based on the $N$ particle positions $\{\mathbf{r}_i\}$, that is,
\begin{align}
 g(R)=\frac{2}{\rho\cdot (N-1)\cdot 4\pi\cdot R^2} \,\left<   \sum_{i<j \in \{1,\dots,N\}} \delta(\left|\mathbf{r}_j - \mathbf{r}_i\right| - R) \right>\textrm{,}
 \label{eqn:gka}
\end{align}
is designed to not distinguish among particle species.

In the next step, we set out to find a monodisperse isotropic system with the same particle density, whose RDF is equivalent to the one of the KA system.
Moreover, for computational reasons, it would be desired if the total potential energy in that effective system has only pair-wise contributions.
The existence of such underlying pair potential, however, is not proven, 
but as soon as a system with only pair-wise energy contributions with a pair potential
yielding the reference RDF exists,
it is unique according to Henderson.\cite{Henderson1974}
We further want to point out that such a pair potential not only depends on temperature but also on the particle density.
For the interested reader, we recommend the work of Louis\cite{Louis2002} thematizing a discussion concerning density-dependent pair potentials.

In order to obtain the effective pair potential, we chose the iterative Boltzmann inversion (IBI) method.\cite{Soper1996}
The corresponding procedure requires that we simulate at each iteration a monodisperse system with an approximate pair potential $u_i$ and determine an improved pair potential $u_{i+1}$ by making use of the obtained RDF  $g_i$ according to the following iterative formula
\begin{align}
 u_{i+1}(R)=u_i(R)- \alpha_i(R) \,\ln\left(\frac{g_i(R)}{g_{\rm KA}(R)}\right).
 \label{eqn:ibi}
\end{align}
As an initial pair potential, we choose the solution of the hypernetted chain approximation (HNC),\cite{VanLeeuwen1959} that is  
\begin{multline}
 u_0(R)=u_{\rm HNC}(R)=\\=-k_{\rm B} T \,\textrm{ln}(g_{\rm KA}(R)) + k_{\rm B}T \cdot (g_{\rm KA}(R)-c_{\rm KA}(R)-1),
 \label{eqn:uini}
\end{multline}
with $c_{\rm KA}$ being the direct correlation function (implicitly defined through the Ornstein-Zernike equation\cite{Ornstein1914} but explicitly defined in k-space\cite{VanLeeuwen1959}) of the KA system.
If the functions $\alpha_i(R)$ in Eq.~\eqref{eqn:ibi} are carefully chosen, we can reach convergence after a certain iteration index $j$, i.e.
$g_{i\geq j}\approx g_{\rm KA}$.
For this purpose, we implemented in accordance with the original literature\cite{Soper1996} the following mixing function
\begin{align}
\alpha_i(R)=\alpha_0\cdot k_{\rm B}T \cdot\exp(-R^2/(2\,\kappa^2))
\label{eqn:alphai}
\end{align}
with $k_{\rm B}$ being the Boltzmann constant and the remaining parameters were set to $\alpha_0=0.05$ and $\kappa=4\sigma$. The latter parameters are fixed for all $i$'s and represent the main mixing parameter as well as the effective range.
The mixing parameter was chosen quite low, but its low-valued choice is necessary in contrast to standard IBI applications since we have to dampen condensation and crystallization effects appearing during the iteration in this supercooled fluid.
The overall particle number in all systems was set to 35000, which is rather large compared to the original work of Kob and Andersen from 1994,\cite{Kob1994} but allows, on the one hand, to correctly track the generally long-range character of the RDFs $g_i$ and avoids system size effects on the other.
With respect to the pair forces, a cut-off distance of $5\sigma$ turned out to be sufficient. 
All simulations for each iteration were performed with the Gromacs simulation package 4.6.7 covering $60000$ time steps per iteration, whereas a time step has a length of $0.005$ time units (t.u.) being $\sigma \cdot (m_p/\epsilon)^{1/2}$, i.e. each iteration covers a time interval of $300$ t.u. 
In order to obtain the reference RDF $g_{\rm KA}$ for all considered temperatures, we simulated each bidisperse KA system with even half the time step length and extracted the required positional data from the time interval $[7000~t.u., 15000~t.u.]$.
As initial systems for the bidisperse as well as all initial monodisperse systems, we used random configurations that were quickly equilibrated using the steepest descent method, and in order not to disturb the systems at each iteration too much, we additionally took the last system snapshot of the $i$-th iteration as input for iteration ${i+1}$.
Considering the temperature control, we used the Nosé-Hoover thermostat with a time constant of $1$ t.u.

\begin{figure*}
 \begin{center}
  \includegraphics[width=0.95\textwidth]{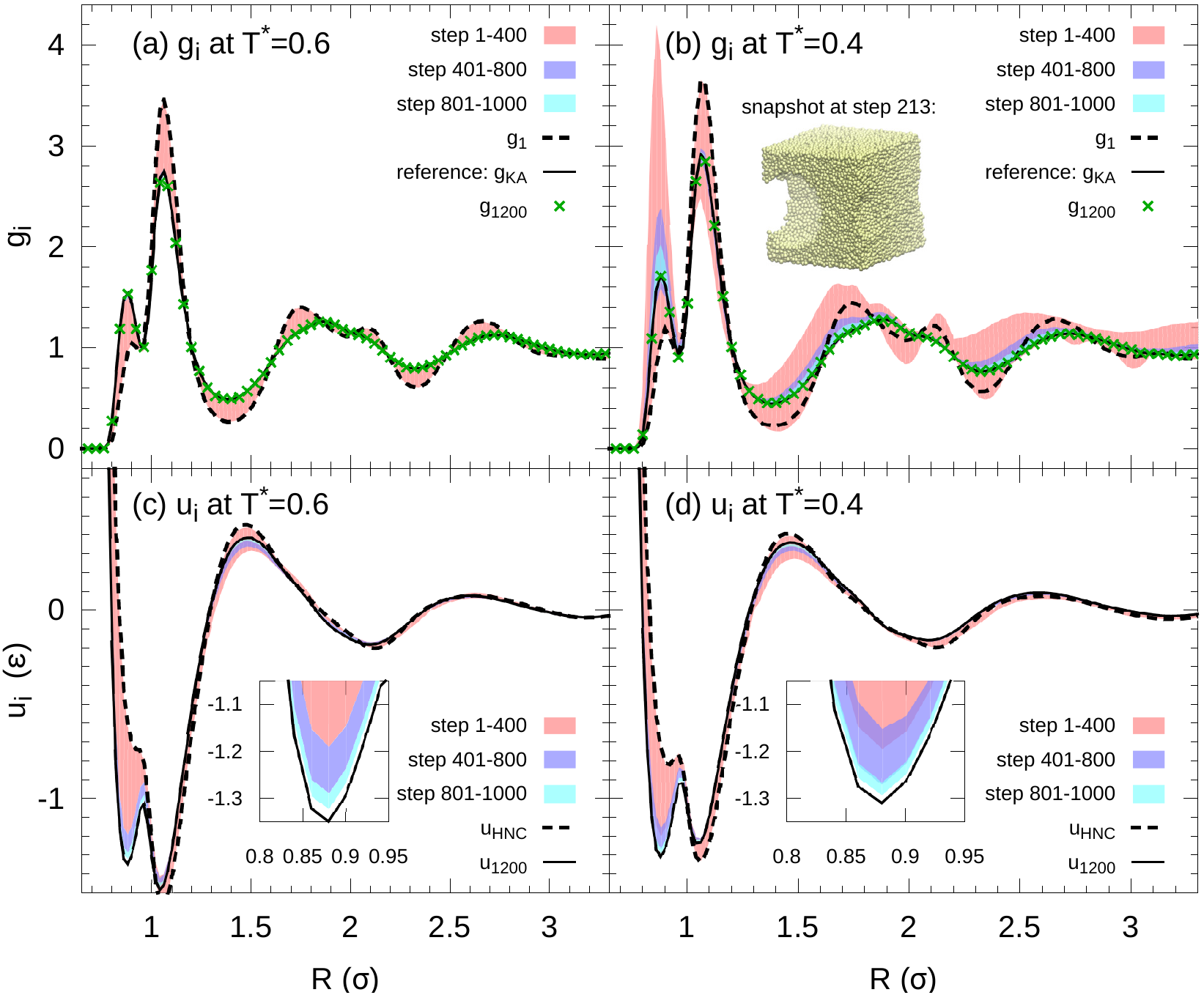}
  \caption{\label{fig:gofruofrkt06and045}Iterative Boltzmann inversion (IBI) results covering 1200 iterations in a 35000 particle Kob-Andersen system for the radial distribution functions (RDFs) $g_i$ (defined through Eq.~\eqref{eqn:gka}) at (a) $T^*=0.6$, (b) $T^*=0.4$ and the corresponding pair potentials $u_i$ at (c) $T^*=0.6$, (d) $T^*=0.4$ are displayed through colored regions.
  Curves in (a, b): the first IBI iteration for the RDF $g_1$  (dashed line); the reference RDF $g_{\rm KA}$ (solid line); final RDF after 1200 steps (green dots).
  Figure (b) additionally contains a snapshot of our system temporarily collapsing at IBI-step 213.
  Curves in (c, d): initial potential $u_{\rm HNC}$ according to Eq.~\eqref{eqn:uini} (dashed line); final result for the effective potential after 1200 iterations (solid line).
  }
 \end{center}
 \end{figure*}

After about 1200 iterations, convergence has been achieved in all our systems in the considered (dimensionless) temperature range covering $T^*=k_{\rm B} T/\epsilon=0.025, 0.05, 0.1, 0.15,\dots , 1$.
At higher temperatures, the bidisperse system is above the freezing point and accordingly out of supercooling.\cite{Pedersen2018}
The total iteration number of 1200 seems to be quite large but its high value stems from the low choice of $\alpha_0$.
The RDFs $g_i$ are obtained using the last third of the positional data of each iteration run. Though this only covers $100$ t.u., it leads to the convergence of the IBI scheme as exemplary shown in Fig.~\ref{fig:gofruofrkt06and045} 
for the $g_i$ and $u_i$ at temperatures being with (a) $T^*=0.6$ well above and with (b) $T^*=0.4$ below the critical (glass transition) temperature of $T_c^* \approx 0.435$ (see Ref.\,\citenum{Kob1994})
within the framework of the mode-coupling theory (MCT).\cite{Bengtzelius1984,Leutheusser1984}
%
At both temperatures, the first iteration of the RDF, $g_1$ (being the RDF corresponding to $u_{\rm HNC}$) in Figs.~\ref{fig:gofruofrkt06and045}(a-b) reveals a low-valued first peak (at $R\approx 0.85\sigma$) and a high-valued second peak (at $R\approx 1.05\sigma$) with respect to the specific reference RDF  $g_{\rm KA}$ from the bidisperse KA system.
In spite of this observed similar shape characteristic for both $g_1$ at these two temperatures, we detect during the first 400 iterations a significant difference in the convergence behavior that can be observed by 
analyzing the corresponding manifold of solutions marked in red.
At  $T^*=0.6$ (Fig.~\ref{fig:gofruofrkt06and045}(a)) that manifold and the series of manifolds reveals a good-natured convergence, which is reflected by their location between $g_1$ and $g_{1200}$, whereas at $T^*=0.4$ (Fig.~\ref{fig:gofruofrkt06and045}(b)) the manifold covering the first 400 iterations
is quite broad and shows a strong tendency to over- and underestimate peak heights.
Such an extreme overestimation of the first peak is observed since the system shows cavitation effects driven by a negative pressure (see snapshot in Fig.~\ref{fig:gofruofrkt06and045}(b)).
However, these effects become less pronounced the more steps we took into account, and as we later see, the converged RDFs all stem from isotropic phases in which the system
cannot overcome critical cavitation within a reasonable simulation time
of each iteration
and no crystallization is present.

Another aspect we recognize during the IBI convergence is that in contrast to the initially fluctuating convergence behavior of the $g_i$ functions at low temperatures, the associated pair potentials $u_i$ show a rather well-behaved convergence at all temperatures as exemplarily shown in Figs.~\ref{fig:gofruofrkt06and045}(c-d).
This is a consequence of our low-valued mixing parameter $\alpha_0$.
Even though the value of $\alpha_0$ appears small, it provides a feedback that is strong enough to prevent even supercooled systems close to the glass transition from quickly crystallizing.
By scrutinizing the deviation between the converged pair potential $u_{1200}$, and the initial HNC pair potential $u_{\rm HNC}$ in Figs.~\ref{fig:gofruofrkt06and045}(c-d), we detect a strong alignment at large inter-particle distances ($R\gg \sigma$) which is a known feature of the HNC approximation.
However, the HNC approximation failed to correctly predict the effective pair potential $u$ at small inter-particle distances $R$. Therefore, the IBI scheme--or perhaps some promising alternative--turned out to be necessary, even though its implementation and convergence in a large simulation system
is involved with respect to computational requirements.
We further would like to point out that the values for $u_i$ below $R=0.8\sigma$ were extrapolated by cumulatively integrating the pair force along $R$, which we have assumed being equal to the force between A and B particles in that close-contact regime.
Regarding the curves' shape of the converged pair potentials $u_{1200}$ from the IBI scheme (Figs.~\ref{fig:gofruofrkt06and045}(c-d)), we can identify an alternating behavior along the inter-particle distance $R$; especially a strong repulsive shoulder appearing at around $R=1.5\,\sigma$.
To have a comparison of this effective pair potential with pair potentials of the original KA system, we have already included $u_{1200}$ from Figs.~\ref{fig:gofruofrkt06and045}(c-d) in Fig.~\ref{fig:uofrkobandersenwithueff}.

All in all, our IBI convergence behavior seems quite weak compared to the original work of Soper,\cite{Soper1996} who used IBI to simplify water models with a mixing parameter of $\alpha_0=1$.
Our reference system, however, is situated in the supercooled phase, which is rather unstable at temperatures slightly above the glass transition temperature.
Even though the system below the glass transition temperature 
appears stable for a very long time (up to minor changes affecting the compressibility as pointed out in the appendix),
approximate solutions during the IBI scheme might not (see Fig.~\ref{fig:gofruofrkt06and045}(b)). Another problematic convergence issue in the IBI scheme is the high particle density which is increasing the likelihood of clumping due to a higher number of next neighbors.

\subsection{\label{sec:Initial assessments of the RDF-matching approach}Initial assessments of the RDF-matching approach}

\begin{figure}
\begin{center}
 \includegraphics[width=0.95\columnwidth]{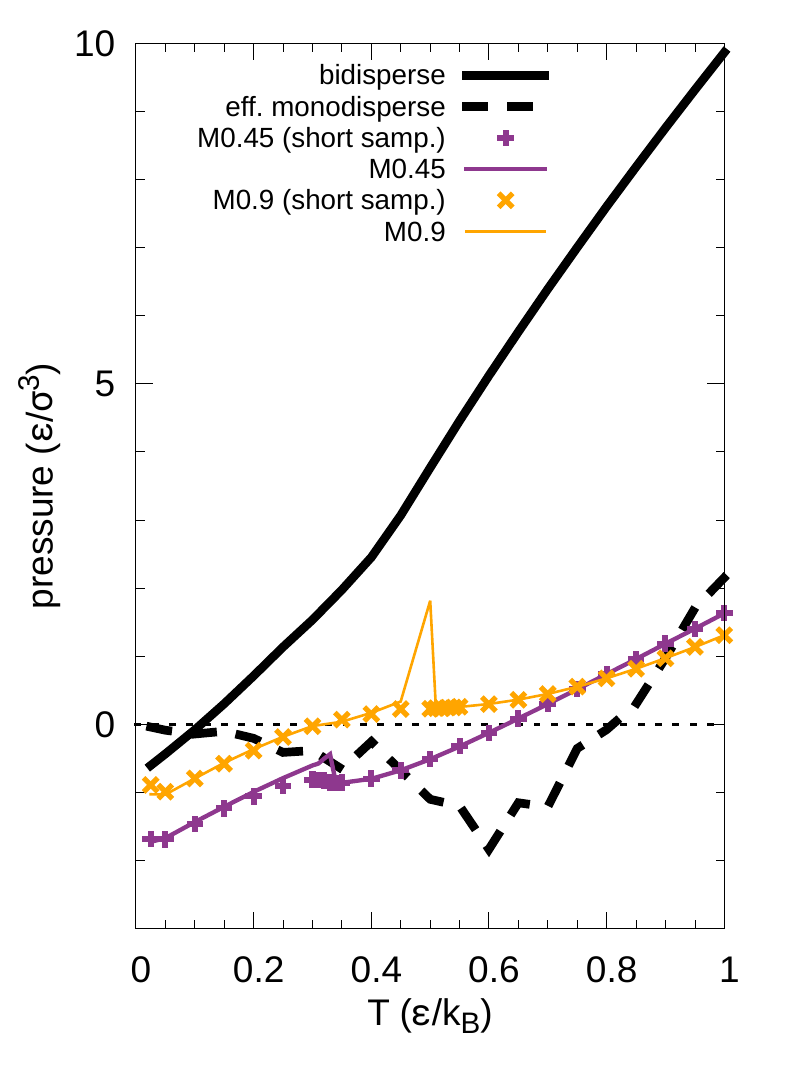}
 \caption{Pressure as a function of temperature covering the original bidisperse Kob-Andersen system, the effective monodisperse system and the later introduced non-intrinsic temperature-dependent monodisperse models M0.45 and M0.9 evaluated using the long sampling scheme (standard) and additionally the short sampling scheme (see Sec.~\ref{sec:Temperature-independent RDF-matching approach}). A horizontal dashed line marks the ${\rm pressure}\!=\!0$ line.
 }
 \label{fig:pressure}
\end{center}
\end{figure}

We start our approach's assessments by investigating the pressure, which is in part also negative for our converged simulations as can be seen through the pressure progression along the temperature in Fig.~\ref{fig:pressure}.
Corresponding ensemble averages (indicated by $\left<\dots\right>$) were ascertained within the framework dubbed as {\it long sampling scheme} comprising an evenly sampling in the time interval $[7000~t.u., 30000~t.u.]$ (forming the standard sampling interval within our investigation).
We recognize that the effective monodisperse system reveals a far lower pressure than the bidisperse system due to less packing fraction 
as a result of a small contact distance which is similar to the one of the A-B configuration (see Fig.~\ref{fig:uofrkobandersenwithueff}).
Effective particles thus can get very close before feeling core repulsion which would significantly increase the pressure in a close packed system.
As a consequence of this volume effect, particles can cluster more easily 
and the phase characterized by the given density becomes meta- or unstable up to a point at which cavitation is unstoppable and the pressure turns negative.

\begin{figure}
\begin{center}
 \includegraphics[width=0.95\columnwidth]{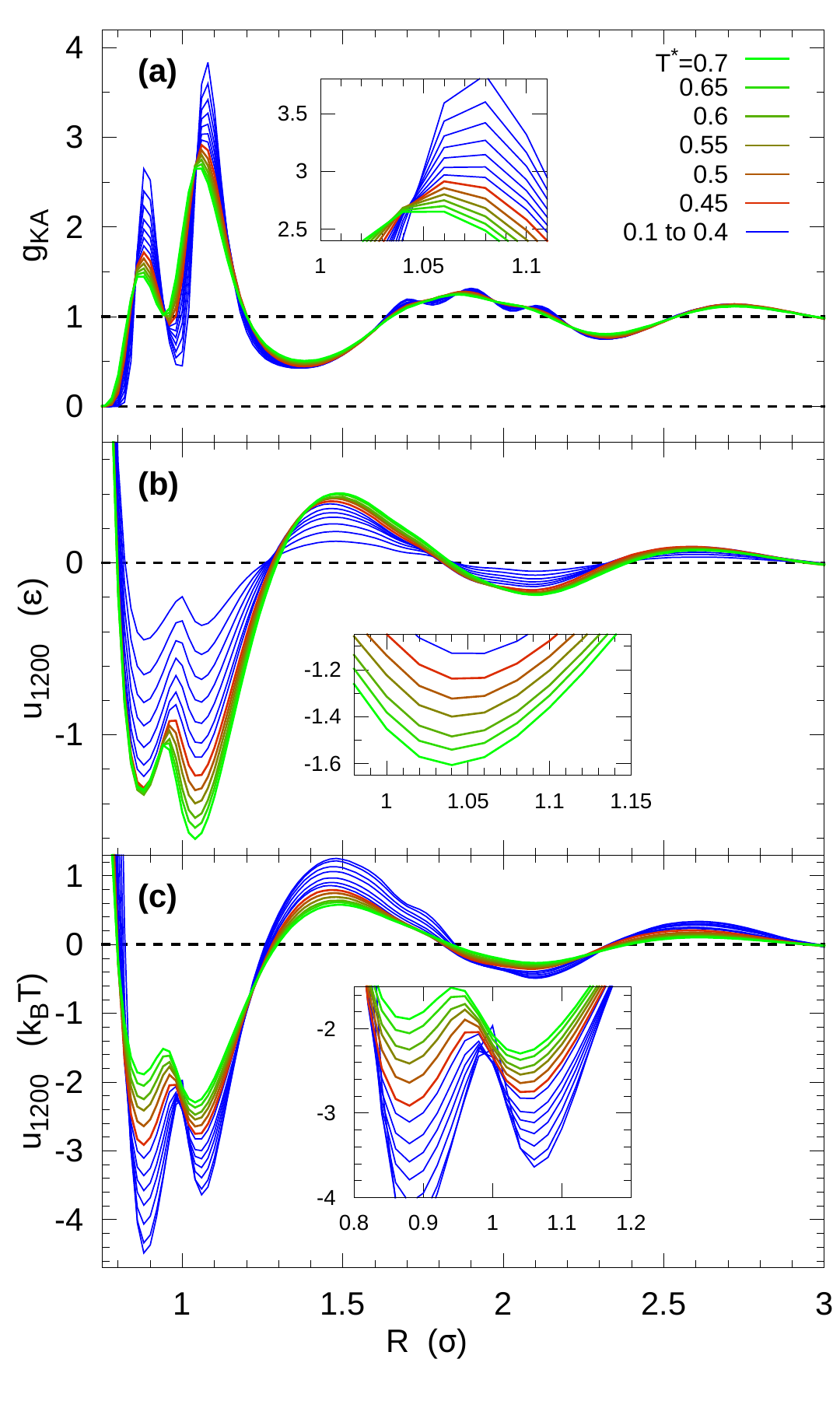}  
\end{center}
 \caption{Results for (a) the RDF in a 35000 particle Kob-Andersen system at different temperatures (in dimensionless form: $T^*=k_{\rm B} T/\epsilon$) and corresponding pair potentials defined through Eq.~\eqref{eqn:ibi} with $i=1200$ in units of (b) $\epsilon$ and (c) $k_{\rm B}T$.}
 \label{fig:gofrktall}
\end{figure}

\begin{figure}
\begin{center}
 \includegraphics[width=0.95\columnwidth]{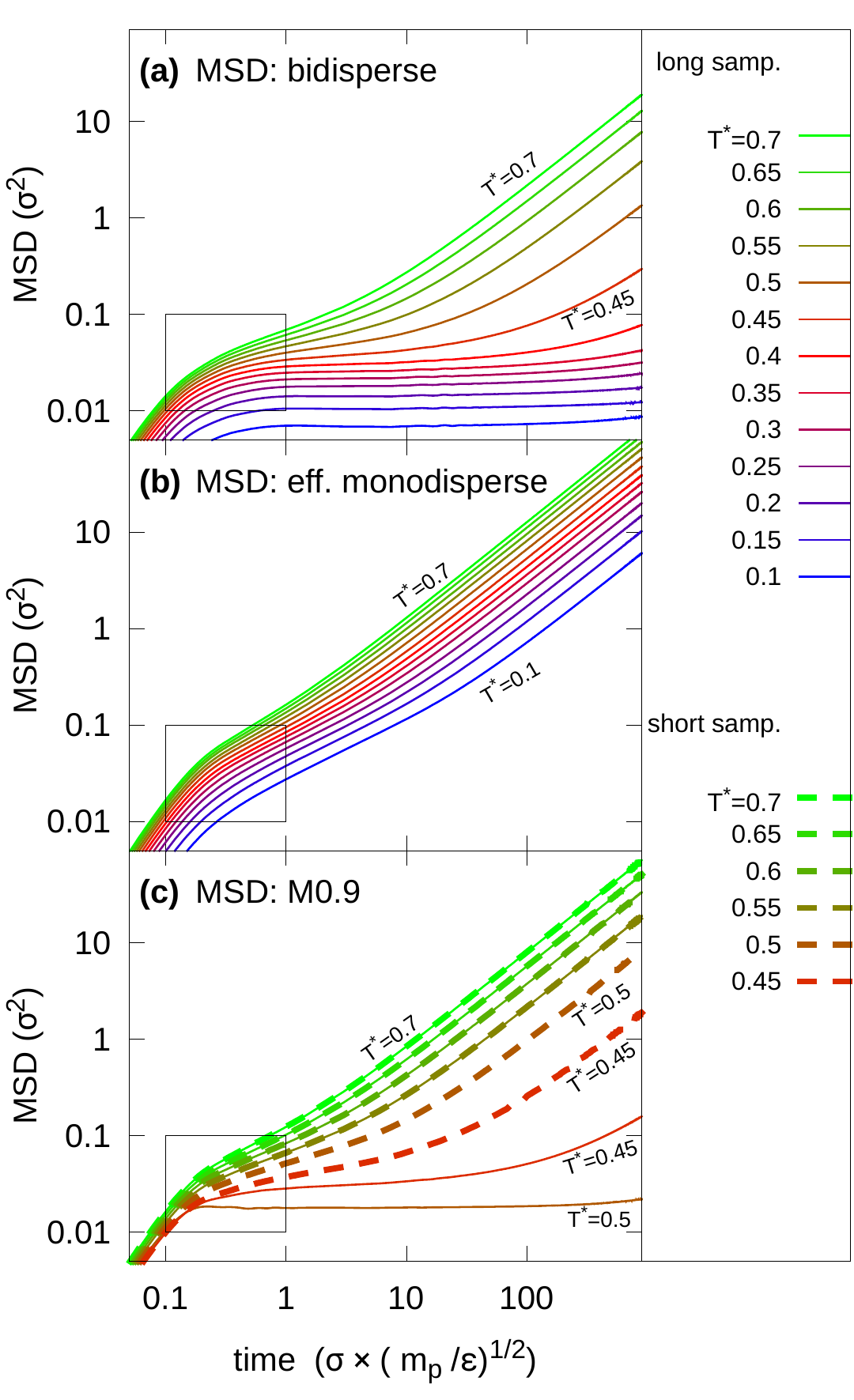}
 \caption{Mean squared displacement for a broad range of temperatures (in dimensionless form: $T^*=k_{\rm B} T/\epsilon$) for (a) the original Kob-Andersen system, (b) the effective monodisperse system and (c) the monodisperse system whose pair interaction is given through the non-intrinsic temperature-dependent M0.9 model.
 Subfig. (c) additionally contains results for a short sampling scheme (for details see main text).
 The rectangles mark a region of interest.
 }
 \label{fig:msd}
\end{center}
\end{figure}

Our next objective is the investigation of the influence of (over- and undercritical) temperature on the reference RDF $g_{\rm KA}$ and its associated converged pair potential $u_{1200}$ (see Fig.~\ref{fig:gofrktall}).
As expected, these possess a slight progression towards the value $1$ (ideal gas limit) the higher the temperature.
But as depicted in the inset, this progression seems to slow down at high temperatures since the system resembles with its steep potentials a hard-sphere mixture before also these pair potentials become softer at higher temperatures.
The associated pair potentials (Fig.~\ref{fig:gofrktall}(b)), however, still reveal a high change rate among neighboring temperature sets.
It is due to the fact that similar RDFs imply a similar Boltzmann weight factor leading to a nearly direct proportionality between $u$ and $T$ in the canonical ensemble average at overcritical temperatures (see inset).
This behavior is in contrast to the temperature dependence of effective pair potentials of particles whose internal degrees of freedom were coarse-grained.\cite{Heinemann2014, Heinemann2015}
As a result, the pair potential in units of $k_{\rm B}T$ (Subfig. (c)) becomes only slightly weaker with temperature and only values at $R \approx 0.9 \sigma$ exhibit stronger weakening effects alongside the temperature progression (see inset).
Given that the effective pair potential is a statistical potential, we interpret this weakening as a consequence of the kinetically-driven higher occurrence of the A-A-particle configuration  with its Lennard-Jones well depth at $R\approx 1.1\sigma$ compared to the A-B configuration with its well depth at $R\approx 0.9\sigma$ (see Fig.~\ref{fig:uofrkobandersenwithueff}).
Our results infer that, since the Boltzmann weight factor is not exactly conserved along temperature, this effective monodisperse pair potential behaves similarly to other works
in the standard ensembles
where CG potentials vary linearly with temperature over a significant temperature range.\cite{farah2011temperature,johnson2007representability,lebold2019systematic,lu2011multiscale}
We now turn the focus also on the dynamics to assess diffusion-related phenomena when transforming from the bidisperse to the effective monodisperse system.
In this regard, we consider the mean squared displacement defined by
\begin{align}
 MSD(t)=\frac{1}{N} \left<\sum_i \left|  \mathbf{r}_i(t)-\mathbf{r}_i(0)  \right|^2  \right>
\end{align}
for the bidisperse and the effective monodisperse system in Fig.~\ref{fig:msd}(a)-(b) for the temperature range $T^*=0.7, \dots,0.1$.
As expected, we detect in the bidisperse system (Fig.~\ref{fig:msd}(a)) a subdiffusive regime (i.e. $MSD\propto t^a$ with $a<1$ holds), where the MSD is almost approaching a constant value when cooling the system. This is reflected by the plateau separating ballistic motion characterized by $MSD(t)\propto t^2$, and diffusive motion, at which $MSD(t)=6Dt$ holds, with $D$ being the diffusion coefficient.
In the effective monodisperse system (Fig.~\ref{fig:msd}(b)), we detect higher MSD-values which results from a lack of a significant subdiffusive regime, in which particles move by collectively rearranging its vicinity, but also from a lower packing fraction. Particles thus undergo collisions later and accordingly leave the ballistic regime on average later what can be seen when comparing the rectangular marked regions in Fig.~\ref{fig:msd}(a,b).
These inconsistencies in dynamics reflect the common failure of structure based coarse-graining.\cite{ghosh2007state,szukalo2020investigation}

\section{\label{sec:Temperature-independent RDF-matching approach}Temperature-independent RDF-matching approach}

In an effort to retain subdiffusive dynamics, we attempt to do the RDF-matching at a reference temperature $T_{\rm ref}$ and intend to use that model --which we denote as $M T_{\rm ref}^*$-model with $T_{\rm ref}^*=k_{\rm B}T_{\rm ref}/\epsilon$ -- for a broad temperature range.
Such a non-intrinsic temperature dependent model is even more desired since also the original bidisperse model has per se no temperature dependence.
Moreover, maintaining microscopic resolution in a temperature-fixed CG model might still result in a good temperature transferability\cite{Izvekov2008} since the ideal gas part of the entropy is not affected.
We chose $T^*_{\rm ref}=0.9$ and started all corresponding simulations from an isotropic configuration for which we took the equilibrated configuration of the effective monodisperse system at $T^*_{\rm ref}$.
In order to asses states with fast crystallization, we also took ensemble averages in a so-called {\it short sampling scheme} covering the early time interval $[100~t.u.,1100~t.u.]$ after starting the simulation. 

\begin{figure}
\begin{center}
 \includegraphics[width=0.95\columnwidth]{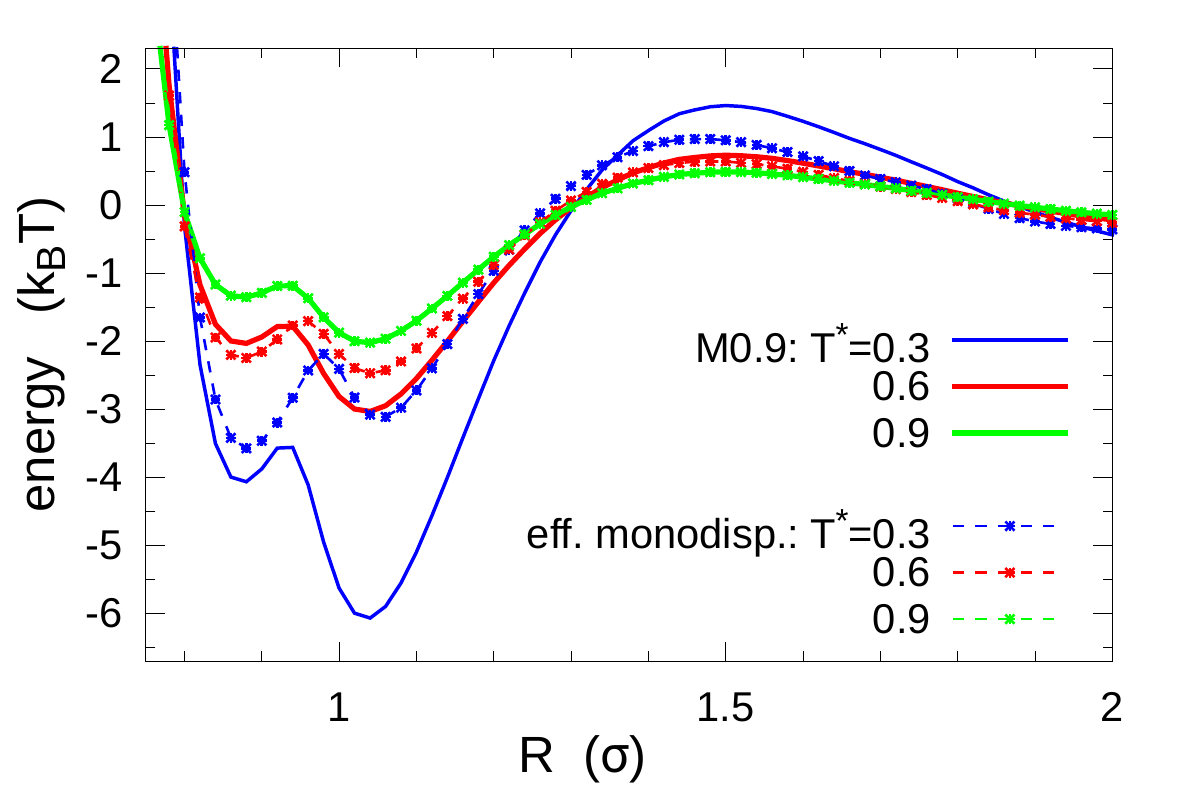}
 \caption{Pair potentials for the M0.9 model (solid lines) and the effective monodisperse model (dashed lines) in units of $k_{\rm B} T$ at the (dimensionless) temperatures $T^*=0.3, 0.6, 0.9$.
 Both models coincide at $T^*=0.9$ by definition.
 }
 \label{fig:m09inkt}
\end{center}
\end{figure}

We observe in a so-formed M0.9-model towards low temperatures for the most part a more dominant pair potential in thermal units compared to those of the effective monodisperse model as depicted in Fig.~\ref{fig:m09inkt}.
This is manifested in more attraction at $R\approx 1.05\sigma$ and more repulsion at $R\approx 1.5\sigma$.
By more attracting direct neighbors but repelling particles slightly further away, we have induced a subdiffusive motion since particles tend to be trapped.
Thus it seems possible to achieve a  glass transition temperature and even tune its value towards the one of the original KA model with the proper choice for $T_{\rm ref}$.
Other works have proposed temperature dependent renormalization approaches of the pair potential for achieving the desired dynamics which also involved corrections of the pair potential towards more cohesive strength at low temperatures.\cite{xia2017energy, xia2018energy, xia2019energy}
These even considered an enlargement of the contact distances (i.e. the packing fraction also increases) which additionally dampens diffusion of IBI models by reducing the excess entropy.\cite{rondina2020predicting,dyre2018perspective}

\begin{figure}
\begin{center}
 \includegraphics[width=0.95\columnwidth]{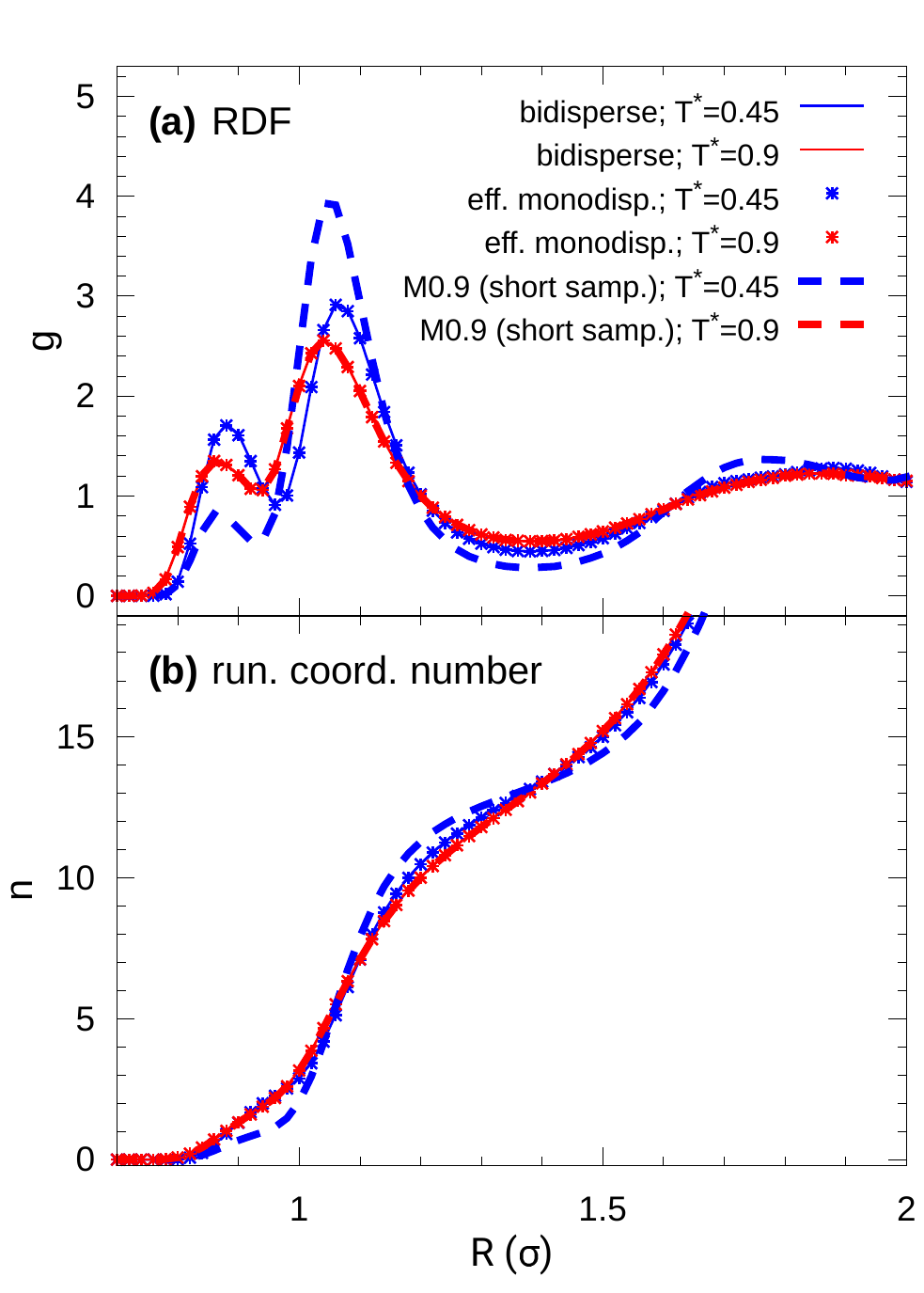}
 \caption{ (a) Radial distribution function, (b) running coordination number of the bidisperse, effective monodisperse, and M0.9-modeled system at two temperatures.
 Results for the latter model are provided within the framework of the short sampling scheme. Results for the effective monodisperse and the bidisperse system match by definition.
 }
 \label{fig:gandruncoor}
\end{center}
\end{figure}

With respect to structure, the M0.9-modeled system possesses at high temperatures the same/similar RDF like the bidisperse system (or equivalently effective monodisperse system) as displayed in Fig.~\ref{fig:gandruncoor}(a).
At low temperature, e.g., $T^*\approx 0.45$, the structural difference becomes more pronounced, what can be seen as an artifact of the increased 
potential well depth at $R\approx 1.05$ leading to a shift of particle distances towards that value.
By extracting from the RDF the running coordination number, defined as
\begin{align}
 n(R)=4\pi\rho \int_0^R dR'\, R'^2 \cdot g(R')
\end{align}
we can recognize that the RDF deviations of M0.9 with respect to the effective monodisperse system along $R$ lead each to an overcompensation of the coordination number deficit or surplus so far.
That means the coordination number differences along $R$ act as diffusive barriers a particle has to overcome and reflects more hampered motion.
As a consequence, MSD results for M0.9 (see Fig.~\ref{fig:msd}(c)) in the short sampling scheme agree with the expected temperature progression of a glass forming fluid (cp. Fig.~\ref{fig:msd}(c) with Fig.~\ref{fig:msd}(a)).
However, the M0.9's MSD temperature progression in the long sampling scheme reveals some abrupt change between $0.55$ and $0.5$ due to crystallization, which
has a faster pace at $T^*=0.5$ (MSD has no diffusive regime) than at $T^*=0.45$.
Fortunately, at lower temperatures, no signs of crystallization have been detected which indicates the glassy state.
Nonetheless, this fast crystallization is contrary to the KA system, which crystallizes at a very long time scale since the one particle species prevents
the other species from crystallizing and these species might even segregate before crystallizing.\cite{Toxvaerd2009} 
So for industrial applications such as hot glue or inks for inkjet printing, which might make use of strong viscosity change around the glass transition temperature, one could avoid the vicinity around glass transition region, e.g., by rapid cooling or heating.

In accordance with the other models, we have also investigated the pressure for the M0.9 (and the M0.45) model in the prior introduced Fig.~\ref{fig:pressure}.
Similar to the (intrinsically temperature-dependent) effective monodisperse model, we also recognize low-pressure profiles in both new models.
Interestingly, the M0.9-modeled system exhibits no negative pressure for temperatures $T^* \gtrsim 0.3$ onwards, irrespective of the used sampling scheme.
We, therefore, observe for this system in the temperature region above its glass transition--which can be estimated from the MSD analysis (Fig.~\ref{fig:msd}(c))--no cavitation.
Contrary to the short sampling scheme we observe in the long sampling scheme, close but above the glass transition temperature, that the pressure curve
reveals for both new models a kink which peaks at about the highest temperature where crystallization has been observed.
At that temperature(s) the dynamics is fast enough to establish a phase transition within the long sampling scheme towards crystallization as pointed out in the forthcoming investigation.

\begin{figure}
\begin{center}
  \includegraphics[width=0.95\columnwidth]{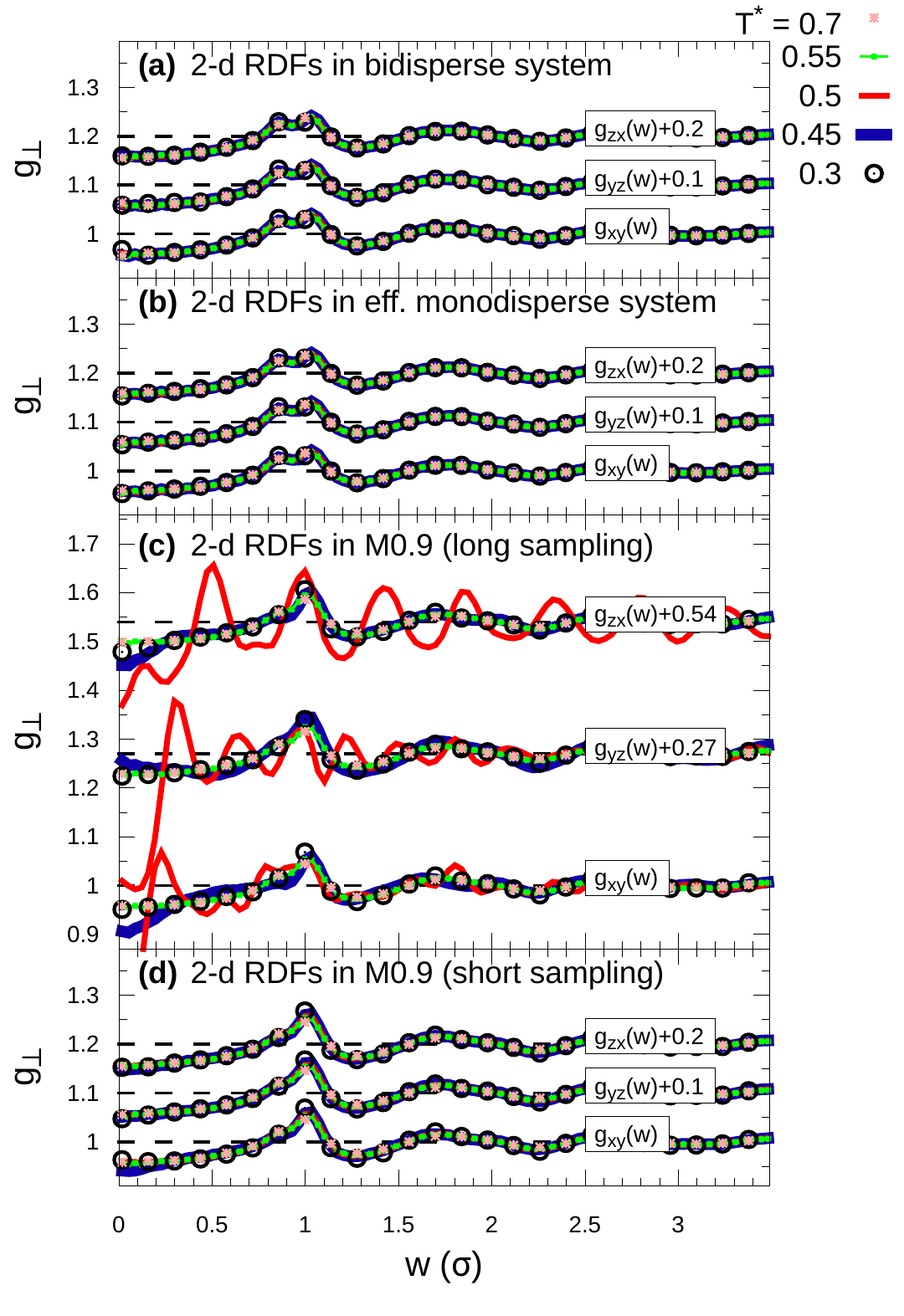}
  \caption{\label{fig:gorthbidispersefixedu} \label{fig:gorthbidispersefixedub} Distribution functions defined by Eqs. \eqref{eqn:gorth1}--\eqref{eqn:gorth4} at various temperatures (in dimensionless form: $T^*=k_{\rm B} T/\epsilon$) covering the range from $T^*=0.7$ with small orange dots to $T^*=0.3$ with large circled black dots (alternatively lines for $T^*=0.45, 0.5, 0.55$) for (a) the bidisperse Kob-Andersen system, (b) the effective monodisperse system and for the M0.9-modeled system for (c) the long, and (d) short sampling scheme (for details see main text).
  }
 \end{center}
\end{figure}

In order to evaluate the influence of using this non-intrinsic temperature-dependent approach on structure, we next investigate for all considered systems two-dimensional RDFs\cite{Andrienko2006} for the particle projections onto the x-y, y-z as well as z-x plane.
These are defined as
\begin{subequations}
\allowdisplaybreaks
\begin{align}
 g_{xy}(w)&=g_{\bot}(w,\mathbf{(0,0,1)}),
 \label{eqn:gorth1}
 \\
  \qquad g_{yz}(w)&=g_{\bot}(w,\mathbf{(1,0,0)}),
  \label{eqn:gorth2}
  \\
  \qquad g_{zx}(w)&=g_{\bot}(w,\mathbf{(0,1,0)}),
 \label{eqn:gorth3}
 \\
 g_{\bot}(w,\mathbf{\hat{n}})\!&=\!\frac{1}{(N-1)\!\cdot\! (N\rho^2)^{1/3}\!\cdot\!\pi w} \!\left<\!\sum_{j, k>j} \delta(w-\left|\mathbf{\hat{n}}\times \mathbf{R}_{jk}\right|)\!\!\right>,
 \label{eqn:gorth4}
\end{align}
\end{subequations}
with $\mathbf{\hat{n}}$ being the normal vector for a considered plane.
In Fig.~\ref{fig:gorthbidispersefixedu}(a-b), we present resulting curves of Eqs. \eqref{eqn:gorth1}--\eqref{eqn:gorth3}
for the bidisperse and the effective monodisperse system at the temperatures $T^* = 0.3, 0.45, 0.5, 0.55, 0.7$. These 2D-RDF curves show neither a dependence on the choice of orthogonal planes nor on the specific model, and only a weak dependence on temperature in accordance with the small temperature dependence of $g_{\rm KA}$ (see Fig.~\ref{fig:gofrktall}(a)). 
As a consequence, the bidisperse and the effective monodisperse systems' structure is isotropic and there is no sign of an abrupt structural change with temperature which would indicate a 
system-wide crystallization. 
For the $M T_{\rm ref}^*$ models, on the contrary, we expect some abnormality in shape since the underlying pair potentials are no longer a function of temperature.
Corresponding distribution functions for the M0.9 model are shown in Fig.~\ref{fig:gorthbidispersefixedub} for the long (c) and the short sampling scheme (d) and reveal that there is
in fact a different structure with respect to all the curves in Subfigs.~(a-b).
Moreover, there is an abrupt structural change with temperature appearing in the long sampling scheme, especially at small projected inter-particle distances $w$.
The shapes at $T^*=0.45, 0.5$ clearly display the anisotropy resulting from the crystallization
by having a pronounced periodic shape, whereas the system remains purely isotropic at shorter time scales  (short sampling scheme used) as depicted in Subfig. (d).
For the latter case, no fundamental structural change can be detected, neither as a function of temperature nor with respect to the plane in which we evaluate $g_{\bot}$, i.e. isotropy remains a meta-stable solution in M0.9 and no bond ordering occurs.
Nonetheless, we detect in Subfig. (d) a deviation from the curves in Subfigs.~(a,b) at $w\approx \sigma$. This results from the fact that both pair potentials only coincide at $T^*=0.9$ and reflects the expected structural change when choosing a non-intrinsically temperature-dependent model.
With respect to the crystallization observed in Subfig. (c) at $T^*=0.45, 0.5$, the structural change  is even higher at $T^*=0.5$ than at $T^*=0.45$, which is an artifact stemming from not enough equilibration time given for completing crystallization,
and as we see in the forthcoming investigation, reflects that this model is close to its glass transition temperature.

\begin{figure}
\begin{center}
 \includegraphics[width=0.95\columnwidth]{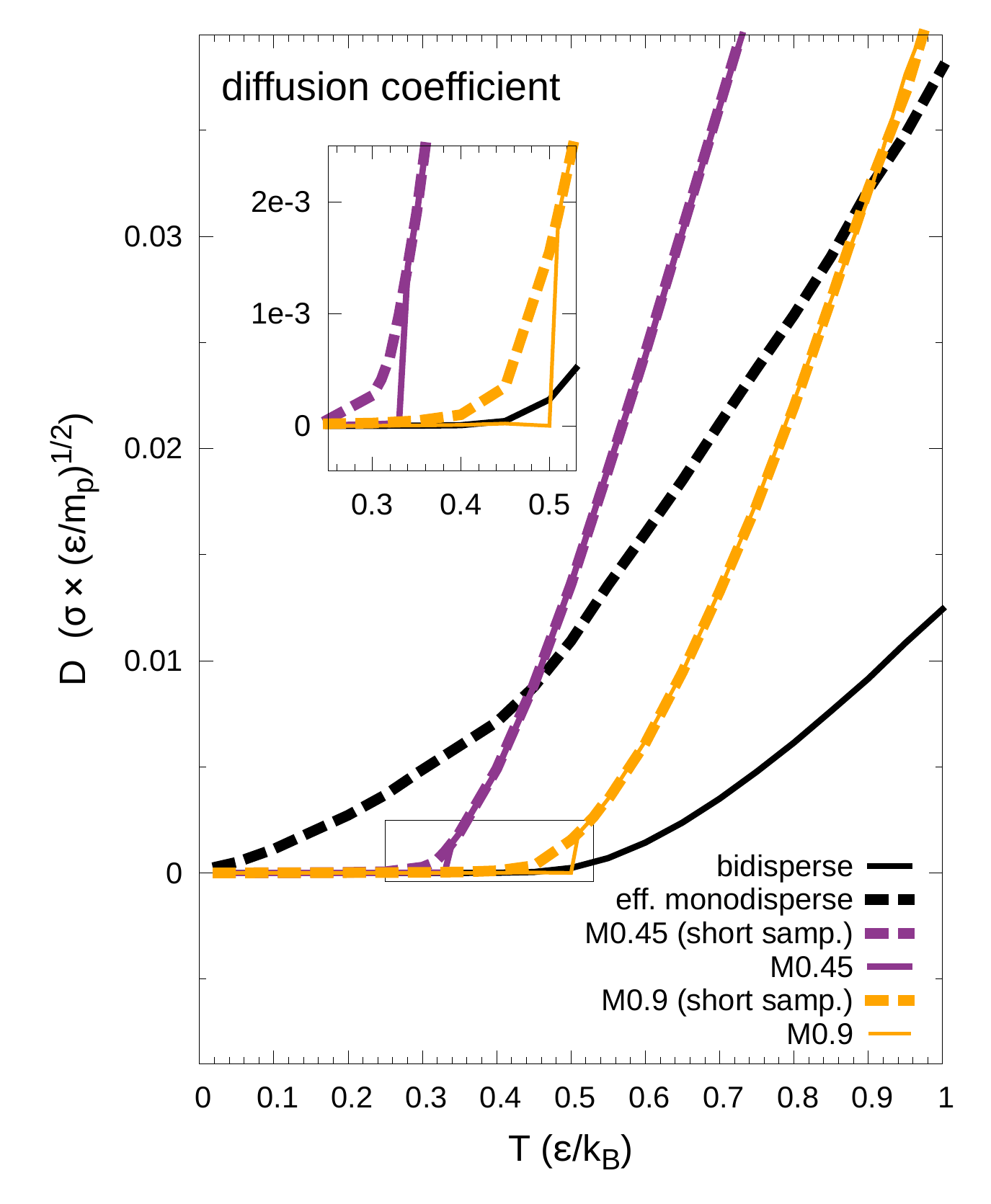}
 \caption{Diffusion coefficient defined by $D=\lim_{t \rightarrow \infty} MSD(t)/6/t$ for the bidisperse Kob-Andersen system (black-solid),  the effective monodisperse system (black-dashed) and for the M0.45 and M0.9-modeled systems. Curves resulting from the short sampling scheme are added for the latter models (colored-dashed).
 The inset magnifies the transition regions between glass and supercooled liquid. Additional simulations in that magnified region have been performed to satisfy a $0.01\, \epsilon / k_{\rm B}$ temperature resolution.
 }
 \label{fig:diffusion}
\end{center}
\end{figure}

\section{\label{sec:Assessing glassy aspects of the systems' dynamics}Assessing glassy aspects of the systems' dynamics}
\subsection{\label{subsec:Diffusion and relxation dynamics}
Diffusion and relaxation dynamics}

This section deals with investigations about the dynamics of former-introduced particle systems in anticipation of detecting glassy aspects. For this purpose, we first present in Fig.~\ref{fig:diffusion} results for the diffusion coefficient $D$ as a function of temperature $T$ and can recognize that $D$ in the bidisperse system follows a path with a monotonous positive slope while having vanishing values below the critical temperature.
The curve of the effective monodisperse system follows also a positive trend but is significantly higher than the bidisperse one and shows no clear inflection point throughout all temperatures.
Consequently, a glass phase cannot be realized.
If the bidisperse as well as the effective monodisperse system's particle distribution were fully described by their equal RDF, 
the systems' excess entropy, $S_{\rm exc}$ would depend exclusively on it and result in equal diffusion which depends exponentially on $S_{\rm exc}$ (see Refs. \citenum{rondina2020predicting, dyre2018perspective}).
The higher packing fraction in the bidisperse system, however, leads to a decrease in the available positions for particles once a number of particle positions is fixed, i.e.
the contribution for $S_{\rm exc}$ caused by higher order correlation functions is smaller (with respect to the effective monodisperse system) and leads to a comparatively low $S_{\rm exc}$ and thus a lower $D$.

Contrary to the effective monodisperse system,
the $M T_{\rm ref}^*$-modeled systems 
reveal an inflection point that can be shifted depending on the choice of $T_{\rm ref}^*$.
By closer inspecting their glass transition region (inset of Fig.~\ref{fig:diffusion}), we denote for ${D\lesssim 2\cdot 10^{-3}\times\sigma \cdot (\epsilon/m_p)^{1/2}}$ a sudden decay of the diffusion, whereas this is not observed in the short sampling scheme (dashed lines).
This phenomenon is consistent with our prior observation of the unusual progression of the MSD in the same temperature region (see Fig.~\ref{fig:msd}(c)) due to the crystallization
leading to positional restraints and accordingly vanishing MSD.
What is also obvious from Fig.~\ref{fig:diffusion} is the fact that $D$ of the  $M T_{\rm ref}^*$-modeled systems grows at a steeper slope above their glass transition temperatures compared to the effective monodisperse system, which is caused by not increasing potential strength (acting as a diffusive barrier) with temperature in the  $M T_{\rm ref}^*$-modeled systems.

A faster crystallization can thus be seen as a side effect of the increase in diffusion and might constitute an advantage or disadvantage depending on the application.
A solution might consist in avoiding these temperatures (by rapid cooling/heating) or by establishing only a moderate increase in diffusion instead.
In particular, one could continuously vary the $T_{\rm ref}$ values at each temperature above the glass transition temperature
but this would further alter the structure since very high values for $T_{\rm ref}$ had to be chosen and would influence the critical exponents of the vitrification.
Moreover, the manufacturing of this temperature dependence might be challenging.
Another approach in preventing crystallization consists in overlaying the pair potential by time and temperature-dependent noise, which does not influence the structure on average.
This, however, would alter the equations of motion, which is not our goal.
Adding a time-independent noise to the pair potential, however, might represent another solution.
The nature of the Newtonian equations of motion would remain untouched but one would further disturb the structure and it would be more difficult to properly simulate the system since 
a very short simulation time step would be required.
Also, manufacturing of particles possessing a fuzzy overlay in their pair-potential might be difficult, if not impossible.

\begin{figure}
\begin{center}
 \includegraphics[width=0.95\columnwidth]{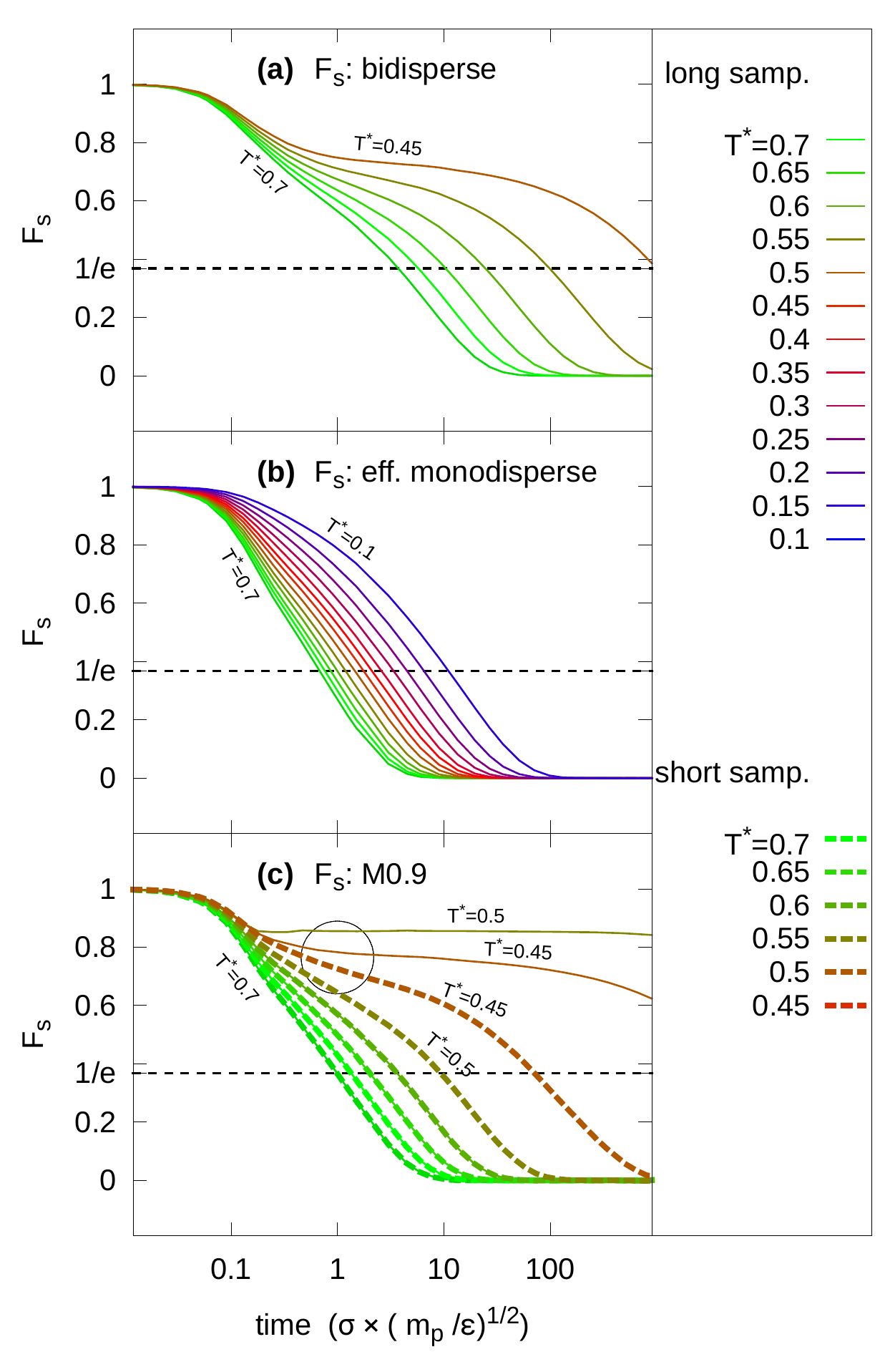}
 \caption{Self-intermediate scattering function defined through Eq.~\eqref{eqn:fs} for a broad range of temperatures (in dimensionless form $T^*=k_{\rm B}T/\epsilon$) for (a) the Kob-Andersen system, (b) the effective monodisperse system and  (c) the M0.9-modeled system covering 35000 particles. Subfig. (c) contains also results for the short sampling scheme (dashed curves).
 The circle marks curves of interest.
 Intersections of the curves with $1/e$ define the $\alpha$-relaxation times.
 }
 \label{fig:fs}
\end{center}
\end{figure}

We next scrutinize the relaxation of the systems in anticipation of an unusual behavior in their subdiffusive regimes.
For this purpose, we present in Fig.~\ref{fig:fs} results for the self-intermediate scattering function, that is
\begin{multline}
 F_s(k,t)=\left<\hat{F}_s(\mathbf{k},t) \right> \text{with}\,\hat{F}_s(\mathbf{k},t)=\\ =\frac{1}{N} \sum_i \cos(\mathbf{k}\circ \left(\mathbf{r}_i(t)-\mathbf{r}_i(0)\right))\text{.} 
  \label{eqn:fs}
\end{multline}
This function characterizes the relaxation of a density mode with wave vector $\mathbf{k}$.
We hereby focus on the spherical component and further set the wavenumber $k$ equal to 
$k_0$, being the wavenumber corresponding to the maximum of the structure factor in the bidisperse system at $T^*=0.45$, yielding $k_0\sigma \approx 7.3$.
As expected, $F_s$ in the bidisperse system (Fig.~\ref{fig:fs}(a)) reveals the characteristic two-step relaxation behavior of supercooled fluids, which was already observed for each particle species separately.\cite{Kim2013}
For the effective monodisperse system (Fig.~\ref{fig:fs}(b)), we only observe a simple decline for all temperatures, which coincides with the absence of a subdiffusive behavior in the MSD (see Fig.~\ref{fig:msd}(b)).
Fortunately, a two-step relaxation dynamics is recovered within our $M T_{\rm ref}^*$-modeled systems as exemplarily shown for the M0.9-modeled system in Fig.~\ref{fig:fs}(c).
In coincidence with the MSD, we also detect slightly below $T^*\lesssim 0.5$ a sudden strong influence of the equilibrium scheme on $F_s$.
In fact, 
we observe in the latter mentioned temperature regime no more alignment between curves stemming from different equilibrium sampling schemes.
We interpret this behavior as an artifact of crystallization (leading to non-vanishing relaxation).
Moreover, $F_s$ at $T^*=0.5$ possesses higher values compared to $F_s$ at $T^*=0.45$, due to a further progressed crystallization process within the long sampling scheme at $T^*=0.5$, which is not finished yet at $T^*=0.45$.
The short sampling scheme, however, leads to a rather ``normal'' temperature progression of $F_s$ (see next paragraph for more details), since the system is not given enough time to crystallize such that it remains supercooled as the bidisperse system.

In order to analyze the temperature progression of $F_s$ qualitatively, we next focus on its decay behavior and thereby quantify the $\alpha$-relaxation regime. Additionally, this also leads us to the characterization of the type of glass-former by focusing on the alpha relaxation time $\tau_{\alpha}$ being implicitly defined through 
\begin{align}
 F_s(\tau_{\alpha})=1/e.
 \label{eqn:taualpha}
\end{align}
\begin{figure*}
\begin{center}
 \includegraphics[width=0.95\textwidth]{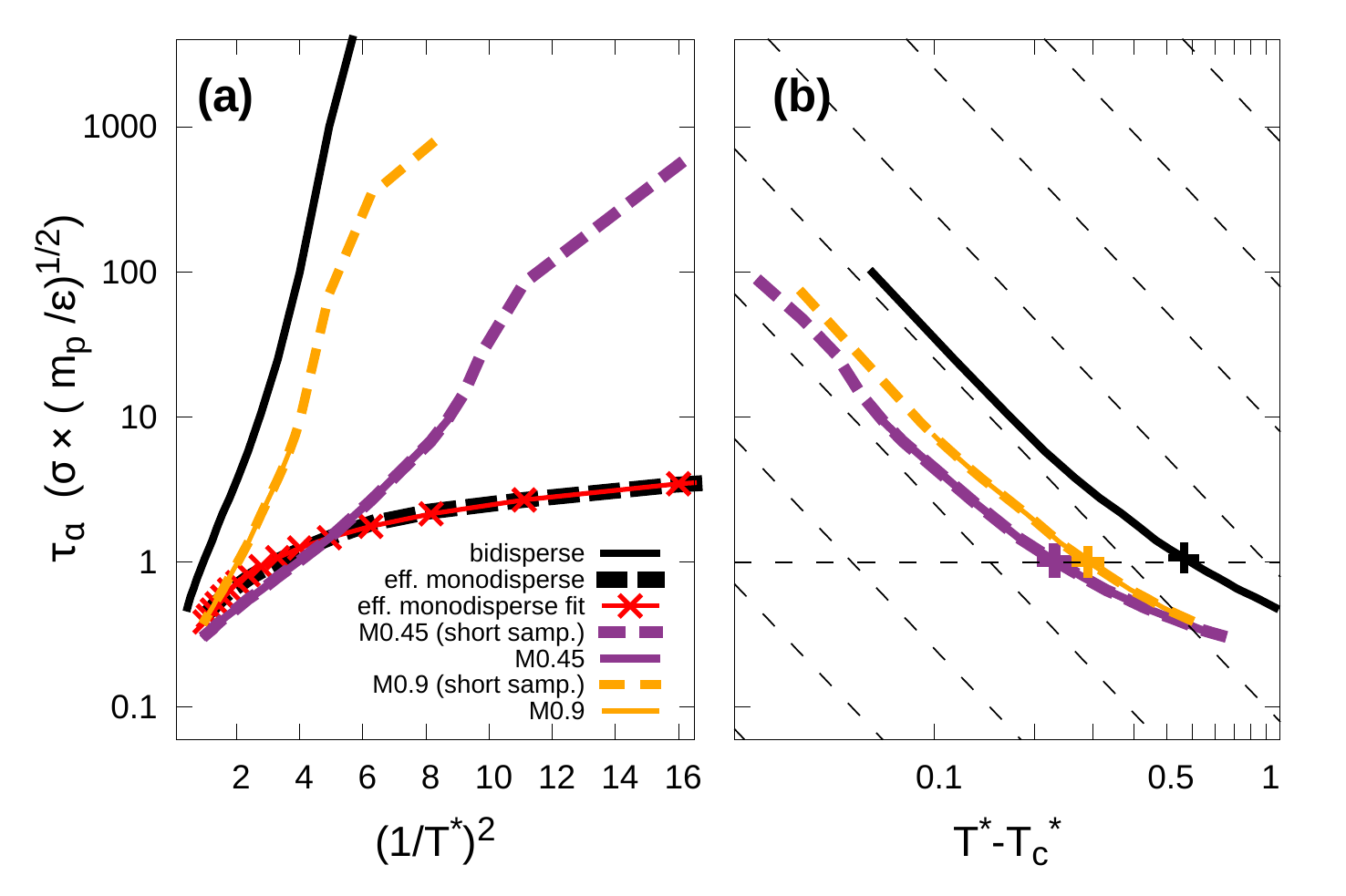}
 \caption{
 (a) Alpha relaxation time $\tau_{\alpha}$ defined through Eq.~\eqref{eqn:taualpha} as a function of the inverse squared temperature (in dimensionless form $T^*=k_{\rm B}T/\epsilon$) for the bidisperse Kob-Andersen system (solid black),  the effective monodisperse system (black-dashed; corresponding fit added as a red curve) and the M0.45 and M0.9-modeled systems (purple; orange). Curves corresponding to a short sampling scheme are added for the latter models (colored-dashed).
 (b) Log-log plot of $\tau_{\alpha}$ as a function of $T^*-T^*_c$ for the bidisperse Kob-Andersen system (with $T_c^*=0.435$), the M0.45-modeled system (with $T_c^*=0.27$), M0.9-modeled system (with $T_c^*=0.41$). The tilted dashed lines represent orientation lines being multiples of 10 of $(T^*-T_c^*) \mapsto (T^*-T_c^*)^{-2.4}$.
 The horizontal dashed line marks the $\tau_{\alpha}$ value of the bidisperse KA system at the melting point.
 }
 \label{fig:fsb}
\end{center}
\end{figure*}
By determining the $\tau_{\alpha}$ values at different temperatures for each model, we can see in Fig.~\ref{fig:fsb}(a) that in the bidisperse system, as well as the M$T^*_{\rm ref}$-modeled systems, the following relation:
\begin{align}
 \tau_{\alpha}\propto \exp[(T^*)^{-p}] 
\label{eqn:tauT}
\end{align}
approximately holds if choosing $p=2$.
This leads to the classification of a fragile glass-former (since $p>1$) and the relaxation behavior is called super-Arrhenius-like.
The same relaxation behavior with $p=2$ is observed for the most part in the M$T^*_{\rm ref}$-modeled systems.
Deviations from super-Arrhenius behavior were observed within the framework of the short sampling scheme and stem from a beginning crystallization in a small temperature interval short above the glass transition temperature (orange and purple dashed lines in Fig.~\ref {fig:fsb}(a)).
Within this interval, the highest deviations occurred inside (rather than at its the edges) where the temperature is low enough and the mobility still high enough to enable initial crystallization during the short observation time.
Results for the long sampling scheme are omitted in that interval due to very large $\tau_{\alpha}$ values which are out of this plotted range.
Concerning the effective monodisperse system, we do not observe a super-Arrhenius law; instead, we could fit a curve (red line) with $p=0.2$.
Accordingly, we yield a sub-Arrhenius behavior for this non-glass-forming model system.
Hereby, the growth in structural relaxation time towards low temperatures is too weak which can be seen as a consequence of the decreasing pair potential strength as shown in Fig.~\ref{fig:gofrktall}(b).

We next provide an explanation for why  $\tau_{\alpha}$ is higher in the M0.9(0.45) compared to the effective monodisperse system by focusing on the Adam-Gibbs theory\cite{adam1965temperature} stating that
\begin{align}
\tau_{\alpha}=\tau_0 \; e^{\frac{\Delta\mu\cdot z}{k_{\rm B}T}}
\label{eqn:AG}
\end{align}
holds with $\tau_0$, $\Delta\mu$ being material constants and $z$ is the number of particles in a hypothetical cooperative spatial region.\cite{dudowiczjcpb2005a,dudowiczjcpb2005b,dudowicz2005jcp,dudowicz2006}
If assuming the same material constants among the M0.9(0.45)-modeled and the effective monodisperse system, the difference has to be primarily in $z$, which
is influenced by the next neighbor structure.
Since the running coordination number in M0.9(0.45) is more wavy at low temperatures compared to the effective monodisperse one (see Fig.~\ref{fig:gandruncoor}(b)), the particles have to overcome barriers which
require the local environment to restructure more and explains the higher $z$ and thus the higher $\tau_{\alpha}$ or lower $D$ accordingly.

In order to quantify the glass transition temperature aside from estimates based on values of the diffusion constant, 
Kob and Anderson\cite{Kob1995} used the power-law fit (as predicted in the mode-coupling theory (MCT))
\begin{align}
 \tau_{\alpha} \propto (T^*-T_c^*)^{-\gamma^{\tau}}
 \label{eqn:pl1}
\end{align}
with $T_c^*$ being the critical temperature and $\gamma^{\tau}$ the corresponding critical exponent.
In Fig.~\ref{fig:fsb}(b) we present the $\tau_{\alpha}$--$(T^*\!-\!T_c^*)$ relationship via a log-log plot for the bidisperse and the M0.9(0.45) model.
With respect to the bidisperse system, we observe with $\gamma \approx 2.4$ when approaching $T_c^*=0.435$ the same  power-law behavior as presented in the work of Kim and Saito\cite{Kim2013} resulting from an $F_s$ defined only among the A-type particles (original work:\cite{Kob1994,Kob1995} $\gamma\approx 2.5(2.6)$ for A(B) particles).
From that plot, we can see that the power-law of Eq.~\eqref{eqn:pl1}
also holds for our M$T^*_{\rm ref}$ models close to their respective glass transition temperature
as well and their exponents exhibit similar values, i.e. $\gamma^{\tau}_{\mathrm{KA}}\approx \gamma^{\tau}_{\mathrm{CG}}$.
In Appendix \ref{app:similargammatau}, we provide reasoning for such similarities.
With respect to the effective monodisperse system, no critical temperature can be extrapolated, which is why it is not displayed.

\subsection{\label{subsec:Stokes-Einstein relation}
{Stokes-Einstein relation}}

We next investigate how the diffusion coefficient $D$ depends on the alpha relaxation time $\tau_{\alpha}$.
If, for example, we consider one large particle that moves slowly with respect to a viscous medium, the diffusion coefficient $D$ can be expressed via the 
Stokes-Einstein (SE) relation, that is 
\begin{align}
 D=\frac{k_{\rm B}T}{6 \pi a \eta}            
 \label{eqn:SE}
\end{align}
with $a$ and $\eta$ being the Stokes radius (the radius of the considered particle) and the shear viscosity, respectively.
Literature devoted to the violation of the SE relation when supercooling a liquid can be found in Refs. \citenum{Chang1994, Kind1992, Ngai2000, Swallen2003, Mallamace2006, Mapes2006, Sengupta2014}.
With respect to $\eta$, Debye provided an explicit expression, that is\cite{Grant1957}
\begin{align}
 \eta=\eta(T,\tau_{\alpha})=\frac{\tau_{\alpha}\,k_{\rm B} T}{ 4\pi a^3}.
 \label{eqn:etadebye}
\end{align}
This expression simplifies Eq.~\eqref{eqn:SE} towards $D\propto \tau_{\alpha}^{-1}$ under the assumption of a fixed (i.e. temperature invariant) Stokes radius $a$.
However, this relationship breaks down when supercooling the fluid.\cite{Yamamoto1998,tarjus1995}
Other works investigating glassy systems\cite{KimSoree2016,ParkSangWon2015,Jeong2010,Jung2004} used a similar but fractional version of this expression, that is
$ D\propto (\tau_{\alpha})^{-\xi}$ to cover the dynamics in the highly supercooled regime.

\begin{figure}
\begin{center}
 \includegraphics[width=0.95\columnwidth]{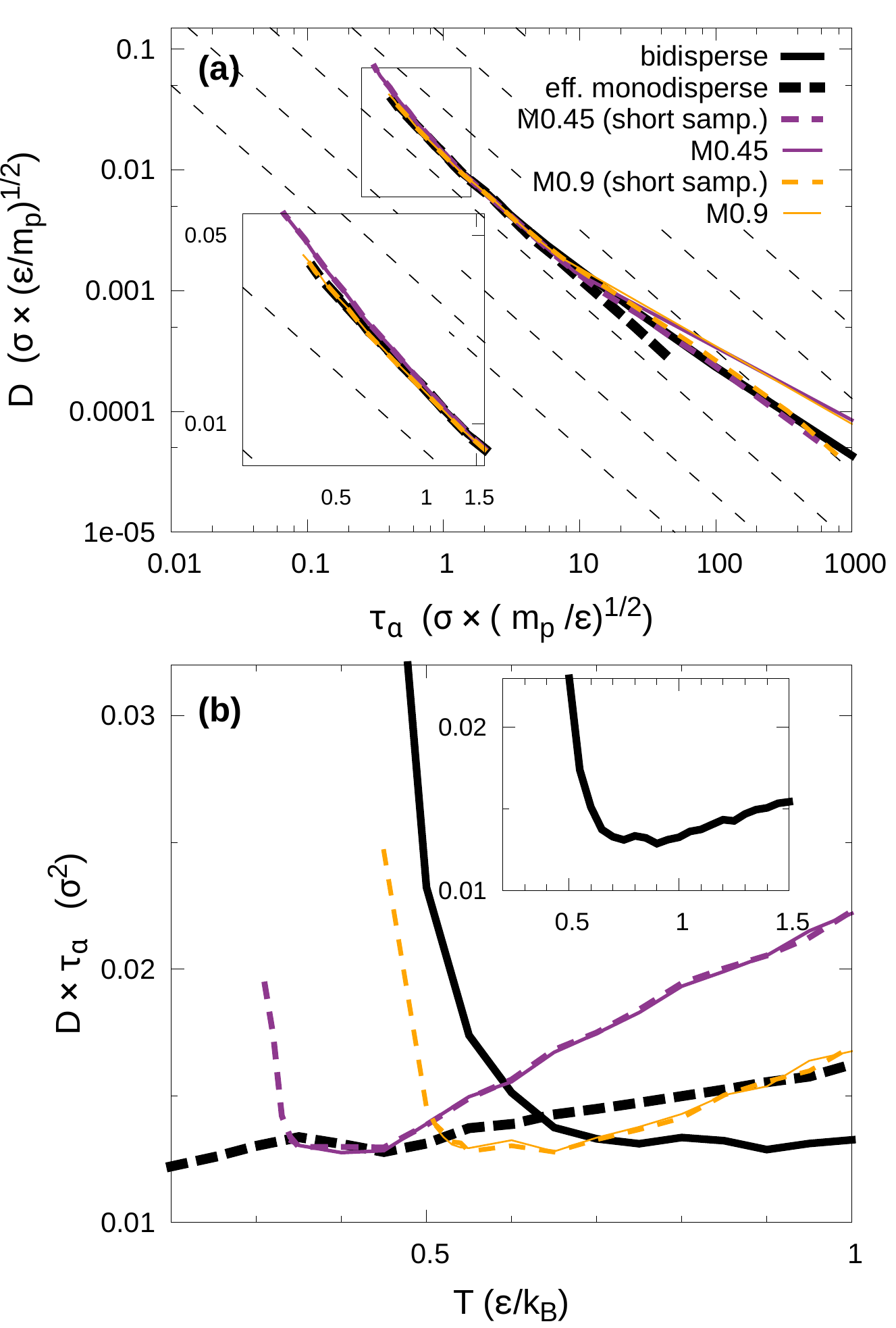}
 \caption{(a) Diffusion coefficient $D$ versus alpha relaxation time $\tau_{\alpha}$. (b) Product of diffusion coefficient $D$ and alpha relaxation time $\tau_{\alpha}$ versus temperature $T$ (Additional simulations for the bidisperse system for $1<T^*\leq 1.5$ have been performed).}
 \label{fig:stokeseinstein}
\end{center}
\end{figure}

In the following, we look for a power-law between $D$ and $\tau_{\alpha}$ for our modeled systems in Fig.~\ref{fig:stokeseinstein}(a) by using a log-log plot.
We find that $\xi$ seems to differ from the ideal value of $1$ for almost all of our models, especially in the limit of a low $\tau_{\alpha}$ (see inset in Fig~\ref{fig:stokeseinstein}(a)) as well as high $\tau_{\alpha}$.
These limits make the curves appear in a slightly convex shape.
With respect to the effective monodisperse system, we detect with $\xi\approx 1.05$ an almost ideal SE behavior, although the glass phase cannot be reached.
In the bidisperse system, however, we yield at large $\tau_{\alpha}$ (i.e. $\tau_{\alpha}^*\gtrsim 10$) or equivalently low $T$ (i.e. $T^*\lesssim 0.6$) with $\xi\approx 0.78$ a value significantly smaller than $1$.
This is expected since we already know from the A- as well as the B-particles' diffusive motion that a break-down of the SE relation towards
a fractional one with $\xi\approx 0.8$ for A-particles, and $\xi\approx 0.65$ for B-particles occurs in the deeply supercooled regime (Ref.~\citenum{Kob1994}).

A further significant point in this power law investigation is the similarity of the $\xi$-values among the systems formed by the CG models M0.9/M0.45 and the bidisperse model. 
In particular, the M0.45/M0.9 systems' $\xi$ coefficients possess in their high temperature regimes with $\xi_{M0.45}(T^*>0.45)=1.35$ and $\xi_{M0.9}(T^*>0.7)=1.22$, respectively, an only slightly higher as well as similar value compared to the one of bidisperse reference system.
For temperatures $T\rightarrow T_c$, the relation $\xi<1$ also holds for our M$T_{\rm ref}^*$-modeled systems with $\xi_{M0.45}(T^*<0.35)=0.79$, $\xi_{M0.9}(T^*<0.6)=0.8$ being close to $0.78$ as detected among all particles in the bidisperse system.
Upon this observation, we provide in Appendix~\ref{app:similargammad} a theoretical explanation by showing similarity in the $\gamma^D=\epsilon\cdot \gamma^{\tau}$ values, i.e. $\gamma^D_{\mathrm{KA}}=\gamma^D_{\mathrm{CG}}$ (similarity for $\gamma^{\tau}$'s already shown in Appendix~\ref{app:similargammatau}).
Therefore, the following relation approximately holds for the bidisperse KA 
and the M$T_{\rm ref}^*$-modeled (CG) systems ($\alpha^D$ values mark scaling parameters).
\begin{align}
 D_{\mathrm{KA}}=\alpha^D_{\mathrm{KA}} (T_{\mathrm{KA}}^*-T_{c,\mathrm{KA}}^*)^{\gamma^D_{\mathrm{KA}}} \propto D_{\mathrm{CG}}=\alpha^D_{\mathrm{CG}} (T_{\mathrm{CG}}^*-T_{c,\mathrm{CG}}^*)^{\gamma^D_{\mathrm{CG}}}.
 \label{eqn:diffprop}
\end{align}
In order to investigate more of the entire $D$-$\tau_{\alpha}$-T relationship, we present in Fig.~\ref{fig:stokeseinstein}(b) a  ``$D \tau_{\alpha}$'' versus ``$T$'' plot,
in which a strong $T$ or $\tau_{\alpha}$ dependence of $D\times\tau_{\alpha}$ at $T^*\lesssim 0.6$ for the bidisperse system  (black-solid line) indicates the observed SE breakdown.
Nonetheless, consistency for that model with the SE relation is found at an intermediate temperature regime ($0.7\lesssim T^*\lesssim 1$) directly beneath the melting point beyond which $\xi=1.19$ holds and 
a linear growth of $D\times\tau_{\alpha}$ with $T$ is found (inset of Fig.~\ref{fig:stokeseinstein}(b)).
Such change in dynamics is also observed in the M$T^*_{\rm ref}$ models at temperatures at which the 
relaxation time is equal to the value observed in the bidisperse system at its melting temperature $T^*\approx 1$ (see Fig.~\ref{fig:fsb}).
The observed linear relationship can be obtained from Eq.~\eqref{eqn:SE} if the shear viscosity $\eta$ is related to $\tau_{\alpha}$ as
\begin{align}
\eta=\eta(\tau_{\alpha})\propto \tau_{\alpha}
\label{eqn:etatau}
\end{align}
and was already observed\cite{Chen2006, Kumar2007} even in fractional form.\cite{Xu2009, Becker2006}
The same dynamic characteristic is also found in our M0.45 and M0.9-modeled systems. 
Investigations involving these two versions of the SE relation (i.e. Eq.~\eqref{eqn:SE} with Eq.~\eqref{eqn:etadebye} or Eq.~\eqref{eqn:etatau}) can be found in Refs. \citenum{Shi2013, Ren2018} whereas
generalized versions of this relation are investigated in Refs. \citenum{Becker2006, Douglas1998}.

\subsection{\label{subsec:statsitical analysis of dynamic heterogeneity}
Statistical analysis of dynamic heterogeneity}
More insights into the dynamics can be obtained by analyzing the particle's activity via excitation events which are characterized by persistence and exchange times defined as follows:\cite{Jeong2010} the persistence time $\tau_p$ for a particle $i$ is the minimal waiting time $t_1$  to undergo its first excitation such
that $\left\|\mathbf{r}_i(t_1)-\mathbf{r}_i(0)\right\|\geq d$ holds whereas
the exchange time $\tau_x$ represents the time between subsequent excitation events, i.e. 
$\left\|\mathbf{r}_i(t_2+t_1)-\mathbf{r}_i(t_1)\right\|\geq d$, $\left\|\mathbf{r}_i(t_3+t_2+t_1)-\mathbf{r}_i(t_1+t_2)\right\|\geq d$, etc.
The parameter $d$ represents the critical displacement for which we chose $d=\sigma$ for all considered models.
A full alignment between the $\tau_p$ and $\tau_x$ distributions means that the probability distribution of the time length, an arbitrarily chosen particle at an arbitrarily chosen simulation time needs to undergo a full excitation event, is equal to those of a particle that just went through such event.
This seems plausible at high temperatures at which there is a very short structural relaxation time $\tau_{\alpha}$ leading quickly to an equal mobility characteristic throughout the system.
If, however,  $\tau_{\alpha}$ becomes more dominant in relation to $\left<\tau_p\right> := \mathbb{E}(P(\tau_p))$, local structures in configuration and momentum space, that have favored a prior excitation event, might still exist and more easily tend to trigger
another event. 
As a consequence of this behavior, the system can exhibit dynamic heterogeneity (spatial partition of the system in dynamically active and inactive regions).

\begin{figure*}
\begin{center}
 \includegraphics[width=0.95\textwidth]{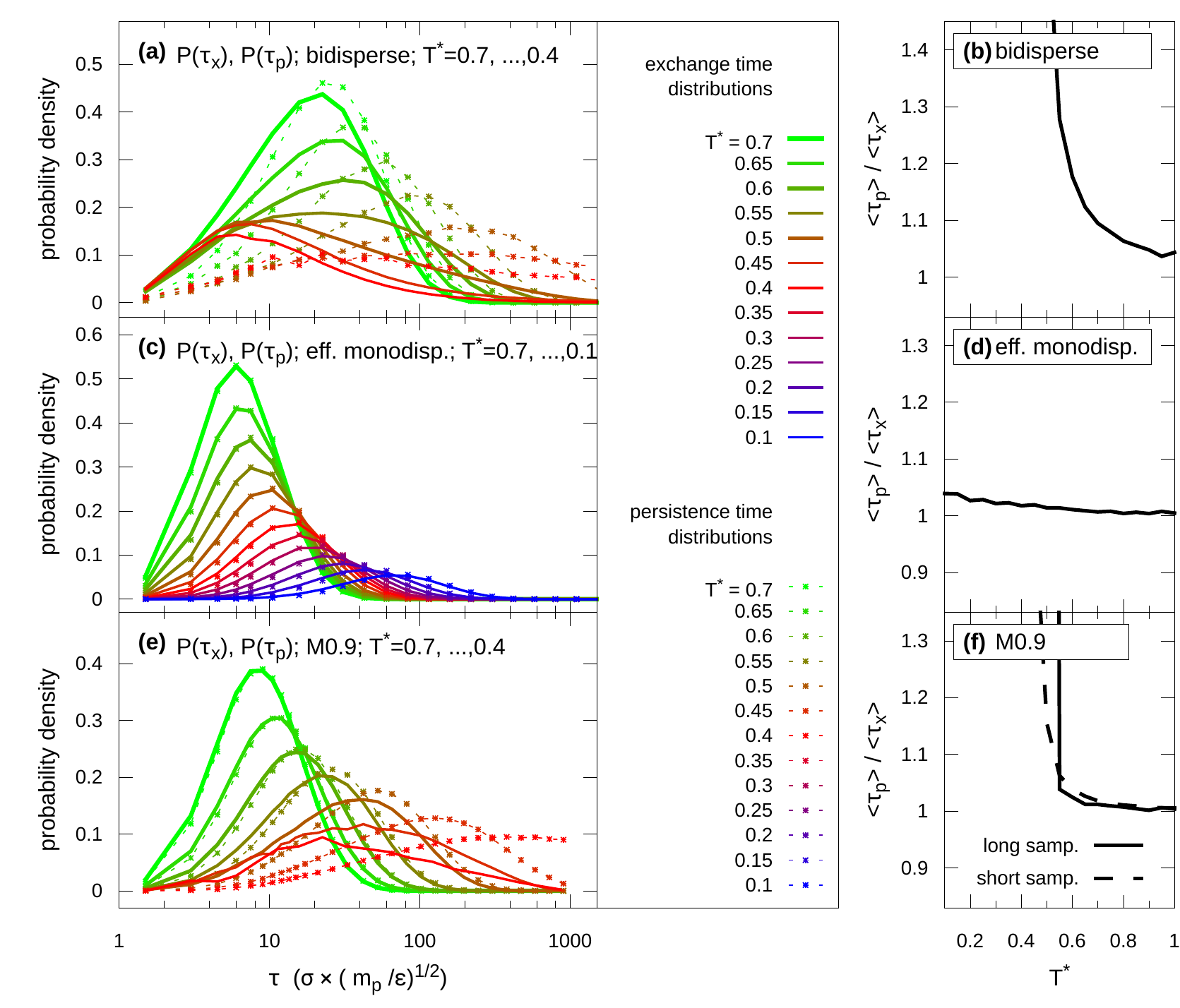}
 \caption{Distributions of the persistence and exchange time and their first moment ratios $\left<\tau_p\right>/\left<\tau_x\right>$  for (a,b) the bidisperse system, (c,d) the effective monodisperse system as well as (e,f) the system modeled via the M0.9 model at various temperatures (in dimensionless form: $T^*=k_{\rm B} T/\epsilon$). For visual simplicity, the distributions were rescaled as a function of temperature. 
 Subfigs.~(a)-(d) display results stemming from the long sampling scheme.
 Subfigs.~(e,f) display results for the short sampling scheme and (f) additionally for the long sampling scheme.
 }
 \label{fig:persistenceandexchangetimesdistributions}
\end{center}
\end{figure*}

In Fig.~\ref{fig:persistenceandexchangetimesdistributions}, the distributions $P(\tau_p)$ and $P(\tau_x)$  and their first moment ratios $\left<\tau_p\right>/\left<\tau_x\right>$ are depicted.
With respect to the bidisperse system (Fig.~\ref{fig:persistenceandexchangetimesdistributions}(a,b)), alignment between $P(\tau_p)$ and $P(\tau_x)$ is found at high temperatures, i.e. $T^* \gtrsim 0.7$.
However, alignment quickly disappears upon cooling leading to persistence times far above excitation times.
As a result, the first moment ratio significantly increases with respect to the ideal value of $1$ upon cooling as depicted in  Subfig.~\ref{fig:persistenceandexchangetimesdistributions}(b).
In the effective monodisperse system (Fig.~\ref{fig:persistenceandexchangetimesdistributions}(c,d)), the distribution alignment holds even at lower temperatures and thus preventing the ratio $\left<\tau_p\right>/\left<\tau_x\right>$ to diverge.
The M0.9-modeled system (Fig.~\ref{fig:persistenceandexchangetimesdistributions}(e,f)), on the contrary, possesses a similar characteristic as those of the bidisperse system as long as we constrain the analysis to the short sampling scheme, at which we have not a progressed crystallization process. In Fig.~\ref{fig:persistenceandexchangetimesdistributions}(f) we provide also the curve resulting from the long sampling scheme.
From that graph, we can see that from  a temperature of $T^*\approx 0.6$ onwards, we have full alignment between both equilibrium sampling schemes, which is expected from our upper investigations.

\begin{figure}
\begin{center}
 \includegraphics[width=0.95\columnwidth]{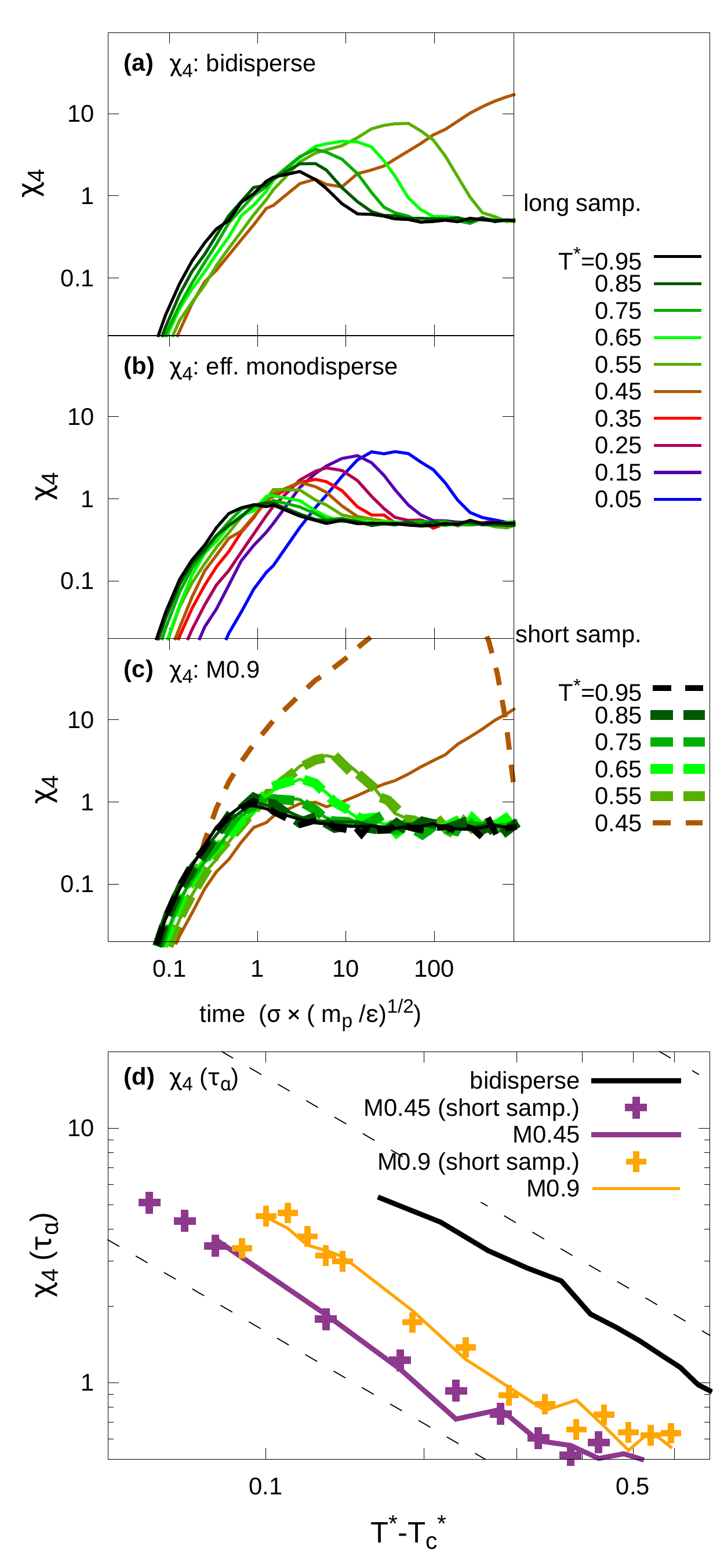}
 \caption{Dynamic susceptibility $\chi_4$ for a broad range of temperatures (in dimensionless form: $T^*=k_{\rm B}T/\epsilon$) for (a) the original Kob-Andersen system, (b) the effective monodisperse system and (c) the M0.9-modeled system. Subfig. (c) contains also results for the short sampling scheme (dashed curves).
 (d) Dynamic susceptibility at the $\alpha$-relaxation time, denoted as $\chi_4(\tau_{\alpha})$, as a function of $T^*-T^*_c$ for the bidisperse Kob-Andersen system (with $T_c^*=0.435$) as well as the M0.45-modeled system (with $T_c^*=0.27$) and the M0.9-modeled system (with $T_c^*=0.41$) for both sampling schemes (solid lines - long sampling scheme; dashed lines/points - short sampling scheme) being defined in the main text. The tilted dashed lines represent orientation lines being multiples of 10 of $(T^*-T_c^*) \mapsto (T^*-T_c^*)^{-1.2}$.
  }
 \label{fig:chi4}
\end{center}
\end{figure}

In order to quantify dynamic heterogeneity for the goal of investigating a corresponding critical exponent, we focus in Fig.~\ref{fig:chi4} on the rescaled variance of the $\hat{F}_s$ function defined in Eq.~\eqref{eqn:fs}, which is also called four-point dynamic susceptibility function,\cite{Toninelli2005} and defined through
\begin{align}
 \chi_4 (k,t)&=N\cdot \left(\left< \hat{F}_s(\mathbf{k},t)^2\right>-\left<\hat{F}_s(\mathbf{k},t) \right>^2\right).
\end{align}
Its maximum value is by definition located where the dynamic fluctuation of $\hat{F}_s$ is maximized and the positions of the maxima follow the alpha relaxation time,\cite{Szamel2006} which we have displayed in Fig.~\ref{fig:fsb}.
With respect to the bidisperse system (Fig.~\ref{fig:chi4}(a)), we observe that the position of the maximum of $\chi_4$ is growing by lowering the temperature.
This is expected since we already know that a similar result was found in a former investigation\cite{Kim2013} of the dynamics of the A-particles in the bidisperse KA system.
In the effective monodisperse system (Fig.~\ref{fig:chi4}(b)), we also observe a growth of the maximum of $\chi_4$ but the maxima of neighboring curves only slightly differ,
meaning that dynamic heterogeneity is not a dominant artifact being observed in this low-fragility ($p=0.2$) fluid while cooling.
This is consistent with a prior investigation,\cite{Kim2013} in which it was observed that more fragility leads to more dynamic heterogeneity.
In Subfig.~(c), we present $\chi_4$-results for the M0.9 model.
Clearly and in a similar fashion to the bidisperse system, we can detect a significant growth of $\chi_4$'s maximum corresponding to strongly increasing dynamic heterogeneity upon cooling
when considering the short sampling scheme.
As expected, we also detect a strong alignment between curves of both sampling schemes at temperatures above the glass transition temperature at which no crystallization occurred within the framework of our
investigation (i.e. $T^* \gtrsim 0.5$).
The growth of $\chi_4$ at the alpha relaxation time, when approaching the critical temperature of the respective model, is further depicted in Fig.~\ref{fig:chi4}(d), where we can roughly recognize in all models exhibiting a glass transition a $\chi_4(\tau_{\alpha})\propto (T^*-T^*_c)^{-\nu}$ behavior with a similar critical exponent of about $\nu\approx 1.2$ as observed for the pure A particle subsystem in Ref. \citenum{Kim2013}.
With respect to the considered temperature regimes in Fig.~\ref{fig:chi4}(d), we restrict to values not
too close to the corresponding critical temperature to avoid artifacts, like roughness or crystallization, due to the requirement of a large sampling time for this variance $\chi_4$.
In order to compare with theoretic predictions and understand why we observe in Fig.~\ref{fig:chi4}(d) similar $\nu$ values, we refer to the inhomogeneous MCT,
which predicted $\nu=1$ (Refs. \citenum{Kim2013, Szamel2006, Biroli2004}).
The critical exponents of the diffusion coefficient $D$, alpha-relaxation time $\tau_{\alpha}$ and 4-point susceptibility $\chi_4$ thus seem to coincide
among the coarse-grained M$T_{\rm ref}$-modeled systems and the bidisperse system.

\section{\label{sec:Conclusions}Conclusions}

In this article, we presented a coarse-graining (CG) method for mixtures for achieving
on the one hand similar structure and higher particle diffusion, which is still based on the Newtonian equations of motion, while satisfying similar dynamic features on the other.
In particular, we aimed at conserving the glass transition temperature, but not all dynamic aspects, which would result in a shift of complexity from the 
static modeling towards more complexity in the equations of motion.
We, however, realized our goals by solely focusing on effective pair potentials
and think that such strategy may contribute to real-world and in silico applications as pointed out further below.
The fundamental concept is based on a reduction of the number of particle species instead of particles itself.
Hereby the well-known bidisperse Kob-Andersen mixture\cite{Kob1994} was transformed into an effective monodisperse system at the same particle density and coarse-grained by conserving in a first attempt the all-particle radial distribution function (RDF).

As a result, we yield in comparison to the original Kob-Andersen mixture quite high values for the diffusion coefficient, which is desired as it acts to increase dynamics, however, this coefficient is unfortunately too high at low temperatures (see Fig.~\ref{fig:diffusion}) leading to a complete absence of the glass transition, which is a dynamic feature aimed to conserve within this method.
We did not anticipate an absence, although, we expected a shift in the glass transition temperature (towards lower temperatures) since coarse-graining smoothens the energy landscape and   might thus lead to higher diffusion in formerly glassy temperature regions.
This high diffusion is also reflected in the weak growth of potential strength in thermal units among effective particle pairs towards lower temperatures (see dashed lines in Fig.~\ref{fig:m09inkt}).

In order to reintroduce the ability to appear in a glassy state, we softened our structural constraint of an identical RDF at all temperatures by introducing two effective models having no intrinsic temperature dependence.
These models each correspond to the temperature-dependent effective monodisperse model at a specific dimensionless temperature $T_{\rm ref}^*$ and were denoted with M0.45 and M0.9 (corresponding to $T_{\rm ref}^*=0.45, 0.9$).
By using this type of approach it was possible to obtain a stronger growth of potential strength in thermal units towards lower temperature (see solid lines in Fig.~\ref{fig:m09inkt}) and thus achieving the glass phase.
The stability of the supercooled region, however, has decreased, such that right above the glass transition temperature a shorter lifetime of the system in the isotropic phase results before crystallization sets in.
Nonetheless, the lifetime is still long enough to reach a two-step relaxation behavior for a short amount of time (see Fig.~\ref{fig:fs}(c) displaying the  self-intermediate scattering function for a short and a long sampling interval).
This faster crystallization results from the increased diffusion which compensates low kinetic energies and thus facilitates the process to overcome potential hurdles for crystallization.
However, this is only an issue at temperatures slightly above the glass transition temperature in the supercooled regime, which is of less extension with respect to the temperature range of the supercooled phase of the mixture.
The general nature of the glass phase, however, is not affected by this phenomenon.
We even observed similar dynamical critical exponents.
Furthermore, we identified that the glass transition temperature can be changed by taking the right choice for  $T_{\rm ref}^*$, which for $T_{\rm ref}^*=0.9$ satisfied a critical temperature of about $0.41$ being close to the value $0.435$ of the original work.\cite{Kob1994}

At this point, we would also like to mention that the underlying methodology can be used for reducing an A,B,C,...-species mixture towards a monodisperse one as well.
More complex cases, for which the outcoming fluid is still a mixture, can be treated in a similar fashion, however, more assumptions might be required and would accordingly constitute another topic of research.

Besides studying physical aspects of this method, our findings suggest a possible use for designing or manufacturing a less or even one-component type of colloidal model fluid out of
complex fluids such as paints, lubricants,
inks for jet printing,
or lacquers which then become disordered solids beneath the same critical (glass transition) temperature as the original mixture.
One could also imagine the development of 
hot glue possessing less starch molecule variety which--due to the higher diffusion at warmer temperatures--better lubricates small pores distributed on surfaces. As a consequence, such glue could provide increased adhesion strength between the materials intended to be glued together.
With respect to machinery, we could imagine the further development of improved motor oil of less added adhesions (or less variety among the often not so uniform nanoparticles) which, e.g., better reduces friction exhibited in the gap between
piston and cylinder or even circumvents patents on certain adhesions.
The one-component fluids resulting from prior applications might be even capable to remain in the liquid state for a broad parameter range, i.e. also in regions where an unwanted eutectic phase separation would occur in the corresponding polydisperse system.
However, designing such effective particles might be difficult or even impossible for certain systems since the effective interactions might be too complex for manufacturing the corresponding colloids or nanoparticles.

There are also in silico applications since a faster diffusion leads not only to shorter relaxation times, which are useful to quickly equilibrate a fluid, there is an additional increase in achievable simulation time
since the interaction range might get shorter (but not necessarily at high densities).
An exemplary system for an in silico application is given through ionic liquids which in their simplest form are described through bidisperse fluids possessing glassy dynamics.\cite{Jeong2011}
Hereby, each ionic particle consists of one or more charged atoms or coarse-grained sites.\cite{Jeong2011,Wang2013}
Computer simulations in such a system normally have the drawback that evolving the equations of motion is quite involved due to long-range electrostatic interactions among numerous charges.
Established bottom-up CG methods dealing with this issue propose a reduction of the atomic detail towards a CG site description\cite{Wang2006,Marrink2004} leading to a reduced electrostatic range due to charge compensation effects within and between CG sites.
For the case of only one CG site per ionic particle, however, one cannot group more atoms into larger CG sites anymore. For this purpose, there exists already a CG method in the interaction space which foots on a usual force matching method
 under the constraint that the effective force is created only by particles within a certain range.\cite{Izvekov2008,Shi2008}
That method is thus more advanced than a simple cut-off scheme. 
Our method would also coarse-grain the interaction space, but contrary to the latter one, by introducing symmetries in it, which might effetively account for the electrostatic screening effect.
In a fully mixed ionic liquid, one can thus also obtain a short range pair potential for the electrostatic part, and accordingly reduce computational costs.
However, more important is the fact that one might design a (perhaps manufacturable) monodisperse model imitating glassy aspects of the two-particle ionic counterpart.

Based on the results of this investigation, we could imagine the following directions of study.
One could construct an efficient algorithm to approach $T_{\rm ref}$, which is the most important parameter in our work for
developing an effective monodisperse model with the same glass transition temperature.
We could also imagine studying the possibility of altering or tuning the chemical structure of a polymer to account for structural influences
caused by other similar polymers in a mixture.
From the methodical point of view, it would be interesting to study the combination of this type of coarse-graining with the classical coarse-graining involving particle reduction.
Another idea concerns the reversion of this method, i.e. one could increase the number of species while having a similar structure to achieve
lower diffusion of particles.
Finally, we would like to point out that the investigation, so far, considers only the inter-particle distance as a variable to describe a pair configuration.
Perhaps, one might extend such an investigation towards more complex pair configuration descriptions.\cite{Heinemann2017}

\section*{Data Availability Statement}
The data that support the findings of this study are available from the corresponding author
upon reasonable request.

\begin{acknowledgments}
This work was supported by the BK21 program and the Samsung Science and Technology Foundation under Project Number SSTF-BA1601-11.
Furthermore, we would like to thank Sang-Won Park, Gyehyun Park and Inrok Oh for fruitful discussions.
\end{acknowledgments}

\appendix


\section{\label{app:similargammatau}Similar critical exponents $\gamma^{\tau}_{\mathrm{KA}}\approx \gamma^{\tau}_{\mathrm{CG}}$ in Eq.~\eqref{eqn:pl1}}    
\begin{figure}
\begin{center}
 \includegraphics[width=0.99\columnwidth]{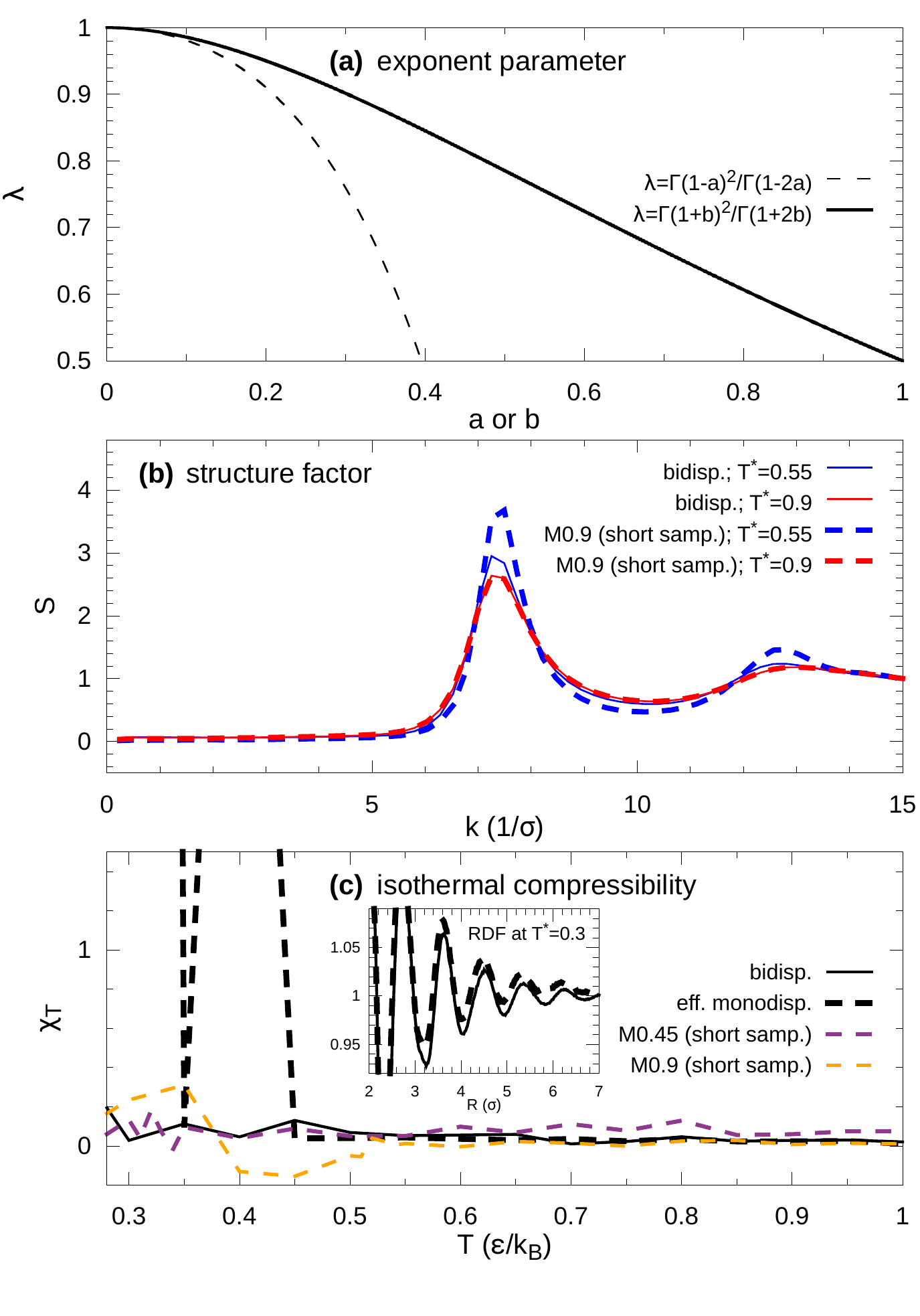}
 \caption{
 The (a) exponent parameter, (b) structure factor, and (c) isothermal compressibility are shown. (b) and (c) cover the bidisperse as well as the M0.9 model, whereas (c) covers also systems modeled via the effective monodisperse and the M0.45 model.
 Results for M0.9(0.45) are provided within the framework of the short sampling scheme.
 }
 \label{fig:gands}
\end{center}
\end{figure}

We present in this section an explanation for why the critical exponent $\gamma^{\tau}$ in the scaling law for the alpha relaxation time (see Eq.~\eqref{eqn:pl1}) reveals similar values for the bidisperse KA and the MT$_{\rm ref}$-modeled CG systems as depicted in Fig.~\ref{fig:fsb}(b).
As a first step, $\gamma^{\tau}$ can  be decomposed into two other coefficients within the framework of the mode coupling theory (MCT), denoted as $a$ and $b$, that is\cite{Kob1995}
\begin{align}
 \gamma^{\tau}&=\frac{1}{2a}+\frac{1}{2b}. 
 \label{eqn:gamma}
\end{align}
These coefficients ($a \in (0,0.4]$ - exponent of critical decay; $b\in (0,1]$ - von Schweidler exponent) can be calculated from the exponent parameter $\lambda$ of the bidisperse system through
\begin{align}
 \lambda=\frac{\Gamma(1-a)^2}{\Gamma(1-2a)}=\frac{\Gamma(1+b)^2}{\Gamma(1+2b)}.
 \label{eqn:lambdabi}
\end{align}
A graphical representation of these parameters is depicted in Fig.~\ref{fig:gands}(a) and reveals the fact that for a given $\lambda$ (with $0.5\leq\lambda<1$), the coefficients $a$ and $b$ can be uniquely determined within the considered range.
The exponent parameter can be approximately related to the structure factor $S$ via\cite{Bengtzelius1984} 
\begin{align}
 \lambda\approx S(k_0)\cdot \frac{k_0\,A^2}{32\pi^2 \rho},
 \label{eqn:lambda}
\end{align}
with $k_0$ and $A$ being the wavenumber corresponding to the major peak in $S$ and its peak area.
In Fig.~\ref{fig:gands}(b) we provide the structure factor $S$ for the bidisperse KA and the CG model M0.9.	
At $T^*=0.9$, $S_{\rm KA}$ and $S_{\rm M0.9}$ coincide per definition (due to the equal RDF), whereas at $T^*=0.55$ (close but not too close to $T^*_c$), the peaks of $S_{\rm M0.9}$ are significantly higher with respect to  $S_{\rm KA}$.
Contrary to the relative growth of $S_{\rm M0.9}$ at $k_0$, we detect a decline around its minima leading to compensation effects for the peak area $A$
and accordingly to a compensation for the growth of $\lambda_{\rm M0.9}$ with respect to $\lambda_{\rm KA}$.
In this respect, we can assume a similar $\lambda$-value resulting in $a_{\rm CG}\approx a_{\rm KA}$ and $b_{\rm CG}\approx b_{\rm KA}$.
Following these thoughts, we finally yield
\begin{align}
\gamma^{\tau}_{\mathrm{KA}}\approx \gamma^{\tau}_{\mathrm{CG}}.
\label{eqn:equalgammas}
\end{align} 
We also want to point out that the upper investigation can be applied to the M0.45 CG model analogously.

\section{\label{app:similargammad}Similar critical exponents $\gamma^D_{\mathrm{CG}}\approx\gamma^D_{\mathrm{KA}}$ in Eq.~\eqref{eqn:diffprop}}    
We next provide a theoretical explanation for the similar critical exponent $\gamma^D$ of the diffusion coefficient (see Eq.~\eqref{eqn:diffprop}) between the bidisperse KA and the CG modeled system M0.9 (M0.45 in a similar way).
Following that investigation and using the connection $\xi=\gamma^D/\gamma^{\tau}$ can explain why our observation from Sec.~\ref{subsec:Stokes-Einstein relation}, stating that $\xi_{\mathrm CG}\approx\xi_{\mathrm KA}$, holds  (similarity for $\gamma^{\tau}$ shown in Appendix~\ref{app:similargammatau}).
For this purpose, we first consider an arbitrary complex fluid system whose dynamics on a large (``diffusive'') time scale is given through the diffusion equation, that is
\begin{align}
\frac{\partial}{\partial t} \rho(\vec{r},t)&=D \Delta \rho(\vec{r},t)
\label{eq:diffeq}
\end{align}
with $\rho(\vec{r},t)$ being the local density (in space and time) in the continuum system or a probability density.
Using this time scale, we can investigate applying the linear response approach, which connects the pressure growth with those of the density in a linear fashion, yielding
\begin{align}
  \rho(\vec{r},t)=\rho_0+(\rho \cdot \chi_T) \cdot p(\vec{r},t).
  \label{eq:linearapproach}
\end{align}
Hereby, $\chi_T$ denotes the corresponding system-wide susceptibility, which is the isothermal compressibility.
As a result of this linear response approach, the diffusion equation \eqref{eq:diffeq} turns into
\begin{align}
\rho \cdot \chi_T \cdot \frac{\partial p}{\partial t}(\vec{r},t) &=D \cdot \Delta \rho(\vec{r},t).
\label{eqn:rhochidpdt}
\end{align}
With this equation we can connect static and dynamic properties since $\chi_T$ can be written in terms of the RDF g through the compressibility equation 
\begin{align}
\chi_T = \frac{1}{\rho\cdot k_{\mathrm B} T} \left[1+\rho \int_0^{\infty} \mathrm{d}R \, 4\pi\,R^2 \, \left(g(R) -1 \right) \right].
\end{align}
The compressibilities of all considered models are displayed in Fig.~\ref{fig:gands}(c) at various temperatures and reveal only slightly positive values above $T^*\approx 0.51$.
The fluids are thus less compressible.
Technically, we would expect a perfect match between curves of the bidisperse and the effective monodisperse model since the compressibility explicitly incorporates the RDF as its only structural information being equal in both models according to the IBI scheme. 
However, there are a few incidences at which the RDF has become rather unstable when equilibrating after our IBI procedure.
We have observed such incidences at $T^*=0.3$ and $T^*=0.4$ at which the compressibility exhibits large positive values due to minor changes in the RDF as exemplarily displayed in the inset in Fig.~\ref{fig:gands}(c) at $T^*=0.3$.

Despite differences close to the glass transition temperature, we can assume a similar compressibility between the M0.9 model (currently referred to as the CG model) and the bidisperse KA model, i.e.
\begin{align}
 \chi_T^{\mathrm{CG} } &\stackrel{!}{\approx}\chi_T^{\mathrm{KA}}.
 \label{eqn:chitreli}
\end{align}
Consequently, Eq.~\eqref{eqn:rhochidpdt} can be applied to the latter identity leading to
\begin{align}
 \chi_T^{\mathrm{CG} }=D_{\mathrm{CG}} \cdot \frac{ \Delta \rho_{\mathrm{CG}}(\vec{r},t)}{\rho\cdot\frac{\partial p_{\mathrm{CG}}}{\partial t}(\vec{r},t)} \approx D_{\mathrm{KA}}  \cdot \frac{\Delta \rho_{\mathrm{KA}}(\vec{r},t)}{\rho\cdot\frac{\partial p_{\mathrm{KA}}}{\partial t}(\vec{r},t)}=\chi_T^{\mathrm{KA}}.
 \label{eq:equalcomp}
\end{align}
We hereby want to point out that we are rather interested in a similarity in their logarithmized values as proposed later. In particular, we mean a similarity even though the computer experiment is not providing a good match.
With respect to the diffusion coefficients $D_{\mathrm{CG}}$ and $D_{\mathrm{KA}}$, it follows from MCT that at temperatures slightly above the corresponding critical temperature $T_c$, the following scaling laws hold:
\begin{subequations}
\allowdisplaybreaks
\begin{align}
D_{\mathrm{CG}}&=\alpha_{\mathrm{CG}} \cdot (T^*_{\mathrm{CG}}-T^*_{c, \mathrm{CG}})^{\gamma^D_{\mathrm{CG}}},
\label{eqn:Da}
\\
D_{\mathrm{KA}}&=\alpha_{\mathrm{KA}} \cdot (T^*_{\mathrm{KA}}-T^*_{c, \mathrm{KA}})^{\gamma^D_{\mathrm{KA}}}.
\label{eqn:Db}
\end{align}
\end{subequations}
Since we chose the temperature $T_{\mathrm{ref}}^*$ in the  MT$_{\rm ref}^*$ CG model M0.9 to satisfy $T_{c, \mathrm{CG}}^*\approx T_{c, \mathrm{KA}}^*=T^*_c$, we 
can focus on the same temperature range $T^*_{\mathrm{CG}}=T^*_{\mathrm{KA}}=T^*>0.51>T^*_c$.
By taking the logarithm of Eq.~\eqref{eq:equalcomp}, which incorporates Eqs.~\eqref{eqn:Da} and \eqref{eqn:Db}, we yield

\begin{widetext}
\begin{align}
 \log\left[\frac{\alpha_{\mathrm{CG}} \Delta \rho_{\mathrm{CG}}(\vec{r},t)}{\rho\cdot\frac{\partial p_{\mathrm{CG}}}{\partial t}(\vec{r},t)}\right]- \log\left[\frac{\alpha_{\mathrm{KA}} \Delta \rho_{\mathrm{KA}}(\vec{r},t)}{\rho\cdot\frac{\partial p_{\mathrm{KA}}}{\partial t}(\vec{r},t)}\right] 
 +\left(\gamma^D_{\mathrm{CG}} -\gamma^D_{\mathrm{KA}}\right) \cdot \log\left( T^*-T^*_c\right) = 0.
\end{align}
\end{widetext}

By, then, applying the limit $T^* \rightarrow [T^*_c]_+$ (and neglecting that the compressibility then slightly turns negative in the M0.9-modeled system close to $T^*_c$), we yield similar critical exponents in the diffusive limit among the bidisperse (KA) and the coarse-grained system, i.e.
\begin{align}
 \gamma^D_{\mathrm{KA}}\approx\gamma^D_{\mathrm{CG}}.
\end{align}
The latter similarity can be motivated for the our other CG model M0.45 as well if assuming for Eq.~\eqref{eqn:chitreli} the expression $\chi_T^{\mathrm{CG} }(T^*) \stackrel{!}{\approx}\chi_T^{\mathrm{KA}}(T^*-T^*_{c, \mathrm{CG}}+T^*_{c, \mathrm{KA}})$.
%

\end{document}